\documentclass[a4paper,12pt]{article}
\usepackage{amssymb}

\usepackage{hyperref}
\usepackage{subfigure}
\usepackage{ae}
\usepackage{mathrsfs}
\usepackage{graphicx}
\usepackage{bbm}
\usepackage{latexsym}
\usepackage{dsfont}
\usepackage{amsmath}
\usepackage[title,titletoc]{appendix}
\usepackage[super]{nth}
\usepackage{cancel}
\usepackage{booktabs}
\usepackage[compat=1.0.0]{tikz-feynman}
\usepackage{multicol}
\usepackage{bbold}

\newcommand{\mathsym}[1]{{}}

\newcommand{\qed}{\nobreak \ifvmode \relax \else \ifdim\lastskip<1.5em \hskip-\lastskip \hskip1.5em plus0em minus0.5em \fi \nobreak \vrule height0.75em width0.5em depth0.25em\fi}

\def\app#1#2{  \mathrel{    \setbox0=\hbox{$#1\sim$}    \setbox2=\hbox{      \rlap{\hbox{$#1\propto$}}      \lower1.1\ht0\box0    }    \raise0.25\ht2\box2  }}

\begin{document}

	\begin{titlepage} 
		\begin{center} \hfill \\
		\hfill \\
			\textbf{\Large
			Decays of the Heavy Top and New Insights on $\epsilon_K$ in a\\ 
				one-VLQ Minimal Solution to the CKM Unitarity Problem}

			\vskip 1cm
			Francisco J. Botella  $^{a,}$\footnote{Francisco.J.Botella@uv.es}, 
			G. C. Branco  $^{b,}$\footnote{gbranco@tecnico.ulisboa.pt}, 
			M. N. Rebelo $^{b,}$\footnote{rebelo@tecnico.ulisboa.pt}, 
			J.I. Silva-Marcos $^{b,}$\footnote{juca@cftp.tecnico.ulisboa.pt} and
			Jos{\'{e}} Filipe Bastos $^{b,}$\footnote{jose.bastos@tecnico.ulisboa.pt} 
			\vskip 0.07in 
			$^a$ Departament de F\' \i sica Te\`orica and IFIC,
			\\
			{\it Universitat de Val\`encia-CSIC, E-46100 Burjassot, Spain.}
			
			$^b$Centro de
			F{\'\i}sica Te\'orica de Part{\'\i}culas, CFTP, \\ Departamento de
			F\'{\i}sica,\\ {\it Instituto Superior T\'ecnico, Universidade de Lisboa, }
			\\ {\it Avenida Rovisco Pais nr. 1, 1049-001 Lisboa, Portugal} \end{center}
		
		\vskip 0.5cm
		\hspace{3.5 cm} PACS numbers :~12.10.Kt, 12.15.Ff, 14.65.Jk 

		\begin{abstract} 
			We propose a minimal extension of the Standard Model where an up-type vector-like
quark, denoted $T$, is introduced and provides a simple solution to the CKM unitarity problem. We adopt the Botella-Chau parametrization in order to extract the $4\times 3$ quark mixing matrix which contains the three angles of the $3\times 3$  CKM matrix plus three new angles denoted $\theta_{14}$, $\theta_{24}$, $\theta_{34}$. It is assumed that the mixing of $T$ with standard quarks is dominated by $\theta_{14}$. Imposing a recently derived, and much more restrictive, upper-bound on the New Physics contributions to $\epsilon _{K}$, we find, in the limit of exact $\theta_{14}$ dominance where the other extra angles vanish, that $\epsilon _{K}^{\text{NP}}$ is too large. However, if one relaxes the exact $\theta_{14}$ dominance limit, there exists a parameter region, where one may obtain $\epsilon _{K}^{\text{NP}}$ in agreement with experiment while maintaining the novel pattern of $T$ decays with the heavy quark decaying predominantly to the light quarks $d$ and $u$. We also find a	reduction in the decay rate of $K_L\rightarrow \pi^0 \nu\overline{\nu}$.
\end{abstract}
	\end{titlepage}
	
\section{Introduction}
	
The normalisation of the first row of $V^{CKM}$ provides one of the most
	stringent tests of $3\times 3$ unitarity of the quark mixing matrix of the
	Standard Model (SM). This results from the fact that the elements $|V_{ud}|$
	and $|V_{us}|$ are measured with high accuracy and $|V_{ub}|$ is known to be
	very small. Recently, new theoretical calculations \cite{Seng:2018yzq} -- 
	\cite{Aoki:2021kgd} of $V_{ud}$ and $V_{us}$ indicate that one may have $%
	\left| V_{ud}\right| ^{2}+\left| V_{us}\right| ^{2}+\left| V_{ub}\right|
	^{2}<1$, thus implying a violation of $3\times 3$ unitarity. If confirmed,
	this would be a major result, providing evidence for New Physics (NP) beyond
	the SM.
	
	It has been pointed out that one of the simplest extensions of the SM which
	can account for this NP, consists of the addition of either one down-type 
	\cite{Belfatto:2019swo} or one up-type \cite{Branco:2021vhs} vector-like
	quark (VLQ) isosinglet. In \cite{Belfatto:2021jhf,Crivellin:2021bkd} both of these possibilities were explored, as well as scenarios with other VLQ representations. In the case of a down-type VLQ isosinglet the CKM matrix consists of the
	first $3 $ rows of a unitary $4\times 4$ matrix, while in the case of an
	up-type VLQ isosinglet, it consists of the first $3$ columns of a $4\times 4$ unitary
	matrix. In both cases, the parameter space is very large, involving six
	mixing angles and three CP violating phases. There are some common features
	in all models with VLQs, such as the appearance of
	Flavour-Changing-Neutral-Currents (FCNC) at tree level \cite{Bento:1991ez}
	-- \cite{Botella:1985gb}. This is a clear violation of the dogma which
	states that no FCNC should exist at tree level. It should be stressed that
	models with VLQs predict the appearance of these dangerous currents, but
	provide a natural mechanism for their suppression. Models with VLQs have a
	rich phenomenology due to the large enhancement of the parameter space.
	
	In this paper, we propose a specific up-type VLQ isosinglet model which solves the
	unitarity problem of the first row of $V^{CKM}$ and makes some striking
	predictions for the dominant decays of the heavy top quark $T$ and for the
	pattern of NP contributions for meson mixings. We adopt the Botella-Chau 
	\cite{Botella:1985gb} parametrization where the new angles are denoted $%
	\theta _{14}$, $\theta _{24}$ and $\theta _{34}$, and assume that $%
	s_{14}\equiv \sin (\theta _{14})$ is the dominant new contribution. In the
	exact $s_{14}$ dominance limit, when $s_{24}=s_{34}=0$, the model predicts:

	\textbf{i)} No tree level contributions to $D^0-\overline{D}^0$ mixing.

	\textbf{ii)} The NP contributions to $B_{d}^0-\overline{B}_{d}^0$ and $B_{s}^0-%
	\overline{B}_{s}^0$ mixings are negligible, when compared to the SM
	contributions.

	\textbf{iii)} The new quark $T$ decays predominantly to the light quarks $d$ and $u$,
	contrary to the usual wisdom.
	
	\textbf{iv)} There are important restrictions arising from NP contributions to $%
	\epsilon _{K}$, specially taking into account the recent results \cite
	{Brod_2012} in constraining the allowed range for NP contributions to $%
	\epsilon _{K}$. In particular, it was shown that it is no longer allowed to
	have a NP contribution to $\epsilon _{K}$ of the same size as the SM
	contribution. In this paper we show that the exact $s_{14}$ dominance limit
	is excluded since it leads to a too large contribution to $\epsilon _{K}$.
	However, we later show that the $s_{14}$ dominance is viable if we allow
	for small but non-vanishing values for $s_{24}$ and $s_{34}$. The
	introduction of small but non-vanishing values for $s_{24}$ and $s_{34}$
	avoids the conflict with $\epsilon _{K}$ while at the same time maintaining
	the distinctive features of the $s_{14}$ limit.
	
	\section{The $s_{14}$ dominance hypothesis: a minimal implementation with
		one up-type VLQ.}
	
	We consider the SM with the minimal addition of one up-type ($Q=+2/3$)
	isosinglet VLQ, denoted by $U_{L}^0$ and $U_{R}^0$.
	
	\subsection{Framework: a minimal extension of the SM with one up-type VLQ}
	
	The relevant part of the Lagrangian, in the flavour basis, contains the
	Yukawa couplings and gauge invariant mass terms for the quarks:
	\begin{equation}
			-\mathcal{L}_{Y}= 
			Y_{u}^{ij}\ \overline{Q}_{Li}^0 \ \tilde{\phi}
			\ u_{Rj}^0+\overline{Y}^{i} \ \overline{Q}_{Li}^0 \ \tilde{\phi}%
			\ U_{R}^0+\overline{M}^{i}\ \overline{U}_{L}^0 \ u_{Ri}^0
			+M\ \overline{U}_{L}^0 \ U_{R}^0 +Y^{ij}_{d} \ \overline{Q}_{Li}^0 \ \phi
			\ d_{Rj}^0+h.c,
		\label{lag}
	\end{equation}
	where $Y_{u,d}$ are the SM up and down quark Yukawa couplings, $\phi $
	denotes the Higgs doublet ($\tilde{\phi}=\epsilon \,\phi ^{*}$), $%
	Q_{Li}^0=\left( u_{Li}^0\,\,d_{Li}^0\right) ^{T}$ are the SM quark
	doublets and $u_{Ri}^0,d_{Ri}^0$ ($i,j=1,2,3$) the up- and down-type SM
	right-handed quark singlets. Here, the $\overline{Y}^{i}$ represent the
	Yukawa couplings to the extra right-handed field $U_{R}^0$, while $%
	\overline{M}$ and $M$ correspond, at this stage to bare mass terms. The
	right-handed VLQ field $U_{R}^0$ is, a priori, indistinguishable from the
	SM fermion singlets $u_{Ri}^0$, since it possess the same quantum numbers.
	
	After the spontaneous breakdown of the electroweak gauge symmetry, the terms
	in Eq. (\ref{lag}) give rise to a $3\times 3$ mass matrix $m=\frac{v}{\sqrt{2%
	}}\,Y_{u}$ and to a $3\times 1$ mass matrix $\overline{m}=\frac{v}{\sqrt{2}}%
	\,\overline{Y}$ for the up-type quarks, with $v\simeq 246$ GeV. Together
	with $\overline{M}$ and $M$, they make up the full $4\times 4$ mass matrix, 
	\begin{equation}
		\mathcal{M}_{u}=\left( 
		\begin{array}{cc}
			m & \overline{m} \\ 
			\overline{M} & M
		\end{array}
		\right)  \label{mu}
	\end{equation}
	One is allowed, without loss of generality, to work in a weak basis (WB)
	where the $3\times 3$ down-quark mass matrix $M_{d}=\frac{v}{\sqrt{2}}%
	\,Y_{d} $ is diagonal, and in what follows we take $M_{d}=D_{d}=\mbox{diag}%
	(m_{d},m_{s},m_{b})$.
	
	The matrix $\mathcal{M}_{u}$ can be diagonalized by a bi-unitary
	transformation 
	\begin{equation}
		\mathcal{V}^{\dagger }\ \mathcal{M}_{u}\ \mathcal{W}=\mathcal{D}_{u}
		\label{du}
	\end{equation}
	with $\mathcal{D}_{u}=\mbox{diag}(m_{u},m_{c},m_{t},m_{T})$, where $m_{T}$
	is the mass of the heavy up-type quark $T$. The unitary rotations $\mathcal{V%
	},\mathcal{W}$ relate the flavour basis to the physical basis.
	
	When one transforms the quark field from the flavour to the physical basis,
	the charged current part of the Lagrangian becomes 
	\begin{equation}
	       \mathcal{L}_{W}=-{\frac{g}{\sqrt{2}}} \overline{u}_{Li}^0 \left( \gamma
		^{\mu }W_{\mu }^{+}\right) d_{Li}^0 =
	       -{\frac{g}{\sqrt{2}}} \overline{u}
		_{L\alpha } \left( \gamma ^{\mu }W_{\mu }^{+}\right) \left( \mathcal{V}%
		^{\dagger }\right)^{\alpha i} d_{Li}
		   \label{char}
	\end{equation}
	where the $u_{L}$ and $d_{L}$ are now in the physical basis. Notice that the
	down quark mass matrix is already diagonal. Thus, we find that the charged
	current quark mixing $\mathcal{V}^{CKM}$ corresponds to the $4\times 3$
	block of the matrix $\mathcal{V}^{\dagger }$ specified in Eq. (\ref{du}) 
	\begin{equation}
		\mathcal{V}^{CKM}=\left( \mathcal{V}^{\dagger }\right) ^{(4\times 3)}
		\label{ckm}
	\end{equation}
	The couplings to the $Z$ boson can be written as
	\begin{equation}
	    \small
		\mathcal{L}_{Z}={\frac{g}{c_W }} Z_{\mu }
		\left[ \frac{1}{2} \left( \overline{u}_{L}F^{u} \gamma ^{\mu }u_{L}-%
		\overline{d}_{L}F^{d} \gamma ^{\mu }d_{L}\right) -s^{2}_W \left( \frac{2}{3} \overline{u}\gamma ^{\mu }u-\frac{1}{3} 
		\overline{d}\gamma ^{\mu }d\right) \right]   \label{Lz}
	\end{equation}
	
	with $F^{d}=\left( \mathcal{V}^{CKM}\right) ^{\dagger }\mathcal{V}^{CKM}$
	and $F^{u}=\mathcal{V}^{CKM}\left( \mathcal{V}^{CKM}\right) ^{\dagger }$. Moreover, one has $c_W=\cos\theta_W$ and $s_W=\sin\theta_W$, where $\theta_W$ is the Weinberg angle.
	
	\subsection{Quark mixing: the Botella-Chau parametrization}
	
	In order to parametrize the $4\times 4$ mixing, we use the Botella-Chau (BC)
	parametrization \cite{Botella:1985gb} of a $4\times 4$ unitary matrix. This
	parametrization can be readily related to the SM usual $3\times 3$ Particle
	Data Group (PDG) parametrization \cite{Zyla:2020zbs} $V^{PDG}$, and is given
	in terms of 6 mixing angles and 3 phases. Defining 
	\[
	V_{4}^{PDG}=\left( 
	\begin{array}{cc}
		\left[ V^{PDG}\right] ^{(3\times 3)} & 
		\begin{array}{c}
			0 \\ 
			0 \\ 
			0
		\end{array}
		\\ 
		\begin{array}{ccccc}
			0 & & 0 & & 0
		\end{array}
		& 1
	\end{array}
	\right) 
	\]
	we can denote the BC parametrization as: 
	\begin{equation}
		\begin{array}{l}
			\mathcal{V}^{\dagger }=O_{34}V_{24}V_{14}\cdot V_{4}^{PDG}= \\ 
			\\ 
			\begin{pmatrix} 1 & 0 & 0 & 0 \\ 0 & 1 & 0 & 0 \\ 0 & 0 & c_{34} & s_{34}
				\\ 0 & 0 & -s_{34} & c_{34}\end{pmatrix}\begin{pmatrix} 1 & 0 & 0 & 0 \\ 0 &
				c_{24} & 0 & s_{24}e^{-i\delta _{24}} \\ 0 & 0 & 1 & 0 \\ 0 &
				-s_{24}e^{i\delta _{24}} & 0 & c_{24}\end{pmatrix}\begin{pmatrix} c_{14} & 0
				& 0 & s_{14}e^{-i\delta _{14}} \\ 0 & 1 & 0 & 0 \\ 0 & 0 & 1 & 0 \\
				-s_{14}e^{i\delta _{14}} & 0 & 0 & c_{14}\end{pmatrix}\cdot \\ 
			\\ 
			\cdot \begin{pmatrix} 1 & 0 & 0 & 0 \\ 0 & c_{23} & s_{23} & 0 \\ 0 &
				-s_{23} & c_{23} & 0 \\ 0 & 0 & 0 & 1\end{pmatrix}\begin{pmatrix} c_{13} & 0
				& s_{13}e^{-i\delta } & 0 \\ 0 & 1 & 0 & 0 \\ -s_{13}e^{i\delta } & 0 &
				c_{13} & 0 \\ 0 & 0 & 0 & 1\end{pmatrix}\begin{pmatrix} c_{12} & s_{12} & 0
				& 0 \\ -s_{12} & c_{12} & 0 & 0 \\ 0 & 0 & 1 & 0 \\ 0 & 0 & 0 &
				1\end{pmatrix}
		\end{array}
		\label{para}
	\end{equation}
	where $c_{ij}=\cos \theta _{ij}$ and $s_{ij}=\sin \theta _{ij}$, with $%
	\theta _{ij}\in [0,\pi /2]$, $\delta _{ij}\in [0,2\pi ]$.
	
	The BC parametrization is such that 
	\begin{equation}
		\left| V_{ud}\right| ^{2}+\left| V_{us}\right| ^{2}+\left| V_{ub}\right|
		^{2}=1-s_{14}^{2},  \label{ckm1}
	\end{equation}
	making it evident that, in this context, a solution for the observed $%
	3\times 3$ CKM unitarity violation implies that the angle $s_{14}\neq 0$.
	
	\subsection{Salient features of $s_{14}-$ dominance}
	
	Let us consider the limit, which we define as the exact $s_{14}$
	dominance, where $s_{14}\neq 0$, while $s_{24}=s_{34}=0$. Then from the
	general Botella-Chau parametrization in Eq. (\ref{para}), and from Eq. (\ref
	{ckm}), we may write for the $4\times 3$ CKM mixing matrix $\mathcal{V}^{CKM}
	$,
	
\begin{equation}
	    \mathcal{V}^{CKM}=
		\left( 
		\begin{array}{ccc}
			c_{12}c_{13}c_{14} & s_{12}c_{13}c_{14} & s_{13}c_{14}e^{-i\delta } \\ 
			-s_{12}c_{23}-e^{i\delta }c_{12}s_{13}s_{23} & c_{12}c_{23}-e^{i\delta
			}s_{12}s_{13}s_{23} & c_{13}s_{23} \\ 
			s_{12}s_{23}-e^{i\delta }c_{12}s_{13}c_{23} & -c_{12}s_{23}-e^{i\delta
			}s_{12}s_{13}c_{23} & c_{13}c_{23} \\ 
			-c_{12}c_{13}s_{14} & -s_{12}c_{13}s_{14} & -s_{13}s_{14}e^{-i\delta }
		\end{array}
		\right) , 
		\label{vs14}
\end{equation}

	where, due to the fact that $s_{24}=s_{34}=0$, the phases $\delta _{24}$ and 
	$\delta _{14}$ may be factored out and absorbed by quark field
	redefinitions. A salient feature of this matrix is that the second and third
	rows of $\mathcal{V}^{CKM}$ exactly coincide with those of the SM $V^{CKM}$.
	In the limit $s_{14}\rightarrow 0$ one recovers the exact SM standard PDG
	parametrization. 
	
	Following \cite{Branco:2021vhs}, we propose here a solution for CKM unitary problem where it is assumed that $s_{14}=O(\lambda^2)$, with $\lambda =|V_{us}|$.
	
	The introduction of vector-like quarks leads to New Physics and consequently
	to new contributions in some very important physical observables. However,
	since in the model considered here with a minimal deviation of the SM
	solving the unitarity problem, one has a mixing where the two angles $%
	s_{24}=s_{34}=0$, some processes, as for instance $D^0-\overline{D}^0$,
	will now have no contributions at tree level. This is also clear from the
	expressions for the Flavour Changing Neutral Currents, where from Eq. (\ref
	{vs14}), one concludes that the FCNC-mixing matrices reduce to 
	\begin{equation}
		\begin{array}{c}
			F^{d}=\left( \mathcal{V}^{CKM}\right) ^{\dagger} \mathcal{V}_{CKM}=\mathbb{1}%
			_{3\times 3} \\ 
			\\ 
			F^{u}=\mathcal{V}^{CKM}\ \left( \mathcal{V}^{CKM}\right) ^{\dagger }=\left( 
			\begin{array}{cccc}
				c_{14}^{2} & 0 & 0 & -s_{14}c_{14} \\ 
				0 & 1 & 0 & 0 \\ 
				0 & 0 & 1 & 0 \\ 
				-s_{14}c_{14} & 0 & 0 & s_{14}^{2}
			\end{array}
			\right) 
		\end{array}
		\label{fcnc}
	\end{equation}
	
	Next, we summarize some of the most salient features of FCNC, in this model:
	
	\textbf{(i)} There is no $D^0-\overline{D}^0$ mixing at tree level, since the $%
	\overline{u}_{L}\ \gamma _{\mu }\ c_{L}\ Z^{\mu }$ coupling does not exist.
	
	\textbf{(ii)} The unique FCNCs at tree level appear in $T\longrightarrow u\ $%
	transitions, coming from the Lagrangian term proportional to $\overline{u}%
	_{L}\ \gamma _{\mu }\ F_{14}^{u}\ T_{L}\ Z^{\mu }$, which leads to the decay 
	$T\longrightarrow u\ Z$, and the term proportional to $\overline{u}_{L}\
	F_{14}^{u}\ T_{R}\ h$ leading to $T\longrightarrow u\ h$:
	
	\begin{figure}[h!]
		\centering
		\begin{multicols}{2}
			\begin{tikzpicture}
				\begin{feynman}
					\vertex (a1){\(u\)};
					\vertex[left=2cm of a1] (a2);
					\vertex[below=1.5cm of a2] (c1);
					\vertex[left=1.5cm of c1] (c2){\(T\)};
					\vertex[below=3cm of a1] (b1) {\(Z\)};
					\vertex[left=0.4cm of c1] (l1);
					\vertex[above=0.1cm of l1] (l2) {\(-s_{14}c_{14}\)};
					
					\diagram* {
						{[edges=fermion]
							(c2) -- (c1) [dot]-- (a1),
						},
						(c1) -- [boson] (b1),
					};
				\end{feynman}
			\end{tikzpicture}

			\begin{tikzpicture}
				\begin{feynman}
					\vertex (a1){\(u\)};
					\vertex[left=2cm of a1] (a2);
					\vertex[below=1.5cm of a2] (c1);
					\vertex[left=1.5cm of c1] (c2){\(T\)};
					\vertex[below=3cm of a1] (b1) {\(h\)};
					\vertex[left=0.4cm of c1] (l1);
					\vertex[above=0.1cm of l1] (l2) {\(-s_{14}c_{14}\)};
					
					\diagram* {
						{[edges=fermion]
							(c2) -- (c1) [dot]-- (a1),
						},
						(c1) -- [scalar] (b1),
					};
				\end{feynman}
			\end{tikzpicture}
			
		\end{multicols}
	\end{figure}
	
	\textbf{(iii)} The charged current couplings of the $T$ quark are
	
	\begin{figure}[h]
		\centering
			\begin{tikzpicture}
				\begin{feynman}
					\vertex (a1){\(d_j\)};
					\vertex[left=2cm of a1] (a2);
					\vertex[below=1.5cm of a2] (c1);
					\vertex[left=1.5cm of c1] (c2){\(T\)};
					\vertex[below=3cm of a1] (b1) {\(W^+\)};
					\vertex[left=0.4cm of c1] (l1);
					\vertex[above=0.1cm of l1] (l2) {\(-t_{14}V_{1j}\)};
					
					\diagram* {
						{[edges=fermion]
							(c2) -- (c1) [dot]-- (a1),
						},
						(c1) -- [boson] (b1),
					};
				\end{feynman}
			\end{tikzpicture}
	\end{figure}
	where $t_{14}\equiv \tan (\theta_{14})$ and the entries $\left( \mathcal{V}^{CKM}\right) _{ij}$ in Eq. (%
	\ref{vs14}) are denoted by $V_{ij}$ with $i,j=1,2,3$ so that $d_j=(d_1,d_2,d_3)=(d,s,b)$.
	
	The most salient feature is the dominant coupling of $T$ to the $d$ and $u$
	quarks and the weakest to the $b$ and top respectively in the channels with $%
	W$ and $Z$ or Higgs. This is quite different from the usual ''wisdom''.
	Experimental bounds on the mass $m_{T}$ of the heavy up-quark are less
	constraining if one does not assume that $T$ quark couples dominantly to $b$
	and top quarks respectively in the decays with $W$ and $Z$ or Higgs.
	
	Let us now consider the new contributions to $B_{d}^0-\overline{B}_{d}^0$, 
	$B_{s}^0-\overline{B}_{s}^0$ mixing, assuming, as stated, that $s_{14}=O(\lambda^2)$ and the known 
	orders in $\lambda$ for the $V_{ij}$.
	
	\subsection*{$B_{d}^0-\overline{B}_{d}^0$ mixing:}

	The NP piece for $B_{d}^0-\overline{B}_{d}^0$ is associated with
	
	\begin{figure}[h]
		\centering
			\begin{tikzpicture}
				\begin{feynman}
					
					\vertex (a1){\(b\)};
					\vertex[right=2.2cm of a1] (a2);
					\vertex[right=1.2cm of a2] (a3){\(W^-\)};
					\vertex[below=2cm of a1] (b1){\(d\)};
					\vertex[right=2.2cm of b1] (b2);
					\vertex[right=1.2cm of b2] (b3){\(W^+\)};
					\vertex[above=0.1cm of a2] (c1){\(t_{14}V_{13}\sim \lambda^5\)};
					\vertex[below=0.1cm of b2] (c2){\(t_{14}V_{11}\sim \lambda^2\)};
					\vertex[below=1cm of a1] (d1);
					\vertex[right=4cm of d1] (d2) {\(\rightarrow \text{ }\lambda^7 \text{  suppression}\)};
					
					\diagram* {
						{[edges=fermion]
							(a1) -- (a2) --[edge label=\(\hspace{1mm}T\)] (b2)--(b1),
						},
						(a2)--[boson] (a3);
						(b2)--[boson] (b3);
					};
				\end{feynman}
			\end{tikzpicture}
    \end{figure}
 while the SM piece is associated with
	\begin{figure}[h]
		\centering		
			\begin{tikzpicture}
				\begin{feynman}
					
					\vertex (a1){\(b\)};
					\vertex[right=2.2cm of a1] (a2);
					\vertex[right=1.2cm of a2] (a3){\(W^-\)};
					\vertex[below=2cm of a1] (b1){\(d\)};
					\vertex[right=2.2cm of b1] (b2);
					\vertex[right=1.2cm of b2] (b3){\(W^+\)};
					\vertex[above=0.1cm of a2] (c1){\(V_{33}\approx 1\)};
					\vertex[below=0.1cm of b2] (c2){\(V_{31}\sim \lambda^3\)};
					\vertex[below=1cm of a1] (d1);
					\vertex[right=4cm of d1] (d2) {\(\rightarrow \text{ }\lambda^3 \text{  suppression}\)};
					
					\diagram* {
						{[edges=fermion]
							(a1) -- (a2) --[edge label=\(\hspace{1mm}t\)] (b2)--(b1),
						},
						(a2)--[boson] (a3);
						(b2)--[boson] (b3);
					};
				\end{feynman}
			\end{tikzpicture}
	\end{figure}
	
so that the dominant contribution to these mixings comes from the SM.
	
	\subsection*{$B_{s}^0-\overline{B}_{s}^0$ mixing:} 
	
	In this case the NP piece is related to
	\newpage
	\begin{figure}[h]
		\centering
			\begin{tikzpicture}
				\begin{feynman}
					\vertex (a1){\(b\)};
					\vertex[right=2.2cm of a1] (a2);
					\vertex[right=1.2cm of a2] (a3){\(W^-\)};
					\vertex[below=2cm of a1] (b1){\(s\)};
					\vertex[right=2.2cm of b1] (b2);
					\vertex[right=1.2cm of b2] (b3){\(W^+\)};
					\vertex[above=0.1cm of a2] (c1){\(s_{14}V_{13}\sim \lambda^5\)};
					\vertex[below=0.1cm of b2] (c2){\(s_{14}V_{12}\sim \lambda^3\)};
					\vertex[below=1cm of a1] (d1);
					\vertex[right=4cm of d1] (d2) {\(\rightarrow \text{ }\lambda^8 \text{  suppression}\)};
					
					\diagram* {
						{[edges=fermion]
							(a1) -- (a2) --[edge label=\(\hspace{1mm}T\)] (b2)--(b1),
						},
						(a2)--[boson] (a3);
						(b2)--[boson] (b3);
					};
				\end{feynman}
			\end{tikzpicture}
	\end{figure}
	whereas for the SM piece one has
	\begin{figure}[h]
		\centering
			\begin{tikzpicture}
				\begin{feynman}
					
					\vertex (a1){\(b\)};
					\vertex[right=2.2cm of a1] (a2);
					\vertex[right=1.2cm of a2] (a3){\(W^-\)};
					\vertex[below=2cm of a1] (b1){\(s\)};
					\vertex[right=2.2cm of b1] (b2);
					\vertex[right=1.2cm of b2] (b3){\(W^+\)};
					\vertex[above=0.1cm of a2] (c1){\(V_{33}\approx 1\)};
					\vertex[below=0.1cm of b2] (c2){\(V_{32}\sim \lambda^2\)};
					\vertex[below=1cm of a1] (d1);
					\vertex[right=4cm of d1] (d2) {\(\rightarrow \text{ }\lambda^2 \text{  suppression}\)};
					
					\diagram* {
						{[edges=fermion]
							(a1) -- (a2) --[edge label=\(\hspace{1mm}t\)] (b2)--(b1),
						},
						(a2)--[boson] (a3);
						(b2)--[boson] (b3);
					};
				\end{feynman}
			\end{tikzpicture}
	\end{figure}
	and again, the dominant contribution arises from the SM.
	
	In the next section, we shall analyse in detail the new contributions to
	some of these physical observables.
	
	\section{Detailed Phenomenological Analysis and New Insights on $\epsilon_{K}$ from Vector-like Quarks}
	
	In this section, we give a more detailed analysis of the previous arguments
	and other new insights, especially focusing on new contributions to $\epsilon _{K}$ from vector-like Quarks.
	
	Because $\mathcal{V}_{42},\mathcal{V}_{43}=0$, there will be no enhancement
	of the rates of rare decays of the top quark into the lighter generations
	(see subsection \ref{NNP}). Therefore, in what follows, we shall focus
	mostly on the contributions to the neutral meson mixings $K^0-\overline{K}%
	^0$ and $B_{d,s}^0-\overline{B}_{d,s}^0$, but we shall also study the
	dominant heavy top decays.
	
	The dominant contributions to some of these processes will depend on $m_{T}$%
	. We restrict our analysis to $m_{T}>685$ GeV, taking into account the CMS
	lower bound for the mass of a heavy top $T$ which couples predominantly to
	the first generation \cite{Sirunyan_2018}.
	
	\subsection{New Physics effects in $K^0-\overline{K}^0$ and $B_{d,s}^0-%
		\overline{B}_{d,s}^0$ mixing}
	
\begin{figure}[h]
		\centering
		\begin{tikzpicture}
			\begin{feynman}
				\vertex (a1) {\(\overline s\)};
				\vertex[right=1.9cm of a1] (a2);
				\vertex[right=1.9cm of a2] (a3);
				\vertex[right=1.6cm of a3] (a4) {\(\overline d\)};
				
				\vertex[below=1.7cm of a1] (b1) {\(d\)};
				\vertex[right=1.9cm of b1] (b2);
				\vertex[right=1.9cm of b2] (b3);
				\vertex[right=1.6cm of b3] (b4) {\(s\)};
				
				\diagram* {
					{[edges=fermion]
						(b1) -- (b2) -- [edge label=\(u_i\)] (a2) -- (a1),
						(a4) -- (a3) --[edge label=\(u_j\)] (b3) -- (b4) ,
					},
					(a2) -- [boson, edge label=\(W\)] (a3),
					(b2) -- [boson, edge label'=\(W\)] (b3)
				};
				
				\draw [decoration={brace}, decorate] (b1.south west) -- (a1.north west)
				node [pos=0.5, left] {\(K^0\)};
				
				\draw [decoration={brace}, decorate] (a4.north east) -- (b4.south east)
				node [pos=0.5, right] {\(\overline{K}^0\)};
			\end{feynman}
		\end{tikzpicture}
		
		\vspace{5mm}
		
		\begin{tikzpicture}
			\begin{feynman}
				\vertex (a1) {\(\overline s\)\hspace{1mm}};
				\vertex[right=1.9cm of a1] (a2);
				\vertex[right=1.9cm of a2] (a3);
				\vertex[right=1.6cm of a3] (a4) {\(\overline d\)};
				
				\vertex[below=1.7cm of a1] (b1) {\(d\)};
				\vertex[right=1.9cm of b1] (b2);
				\vertex[right=1.9cm of b2] (b3);
				\vertex[right=1.6cm of b3] (b4) {\(s\)};
				
				\diagram* {
					{[edges=fermion]
						(b1) -- (b2) --[edge label=\(u_i\)] (b3) -- (b4),
						(a4) -- (a3) --[edge label=\(u_j\)] (a2) -- (a1) ,
					},
					(b2) -- [boson, edge label=\(W\)] (a2),
					(b3) -- [boson, edge label'=\(W\)] (a3)
				};
				
				\draw [decoration={brace}, decorate] (b1.south west) -- (a1.north west)
				node [pos=0.5, left] {\(K^0\)};
				
				\draw [decoration={brace}, decorate] (a4.north east) -- (b4.south east)
				node [pos=0.5, right] {\(\overline{K}^0\)};
			\end{feynman}
		\end{tikzpicture}
		\caption{Leading contributions to $K^0 -\overline{K}^0$ mixing, including
			the effect of the new heavy quark, with $u_{i}=u,c,t,T$.}
		\label{figure1}
	\end{figure}
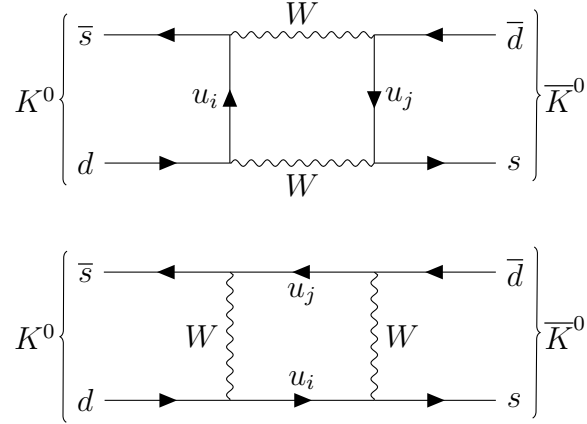

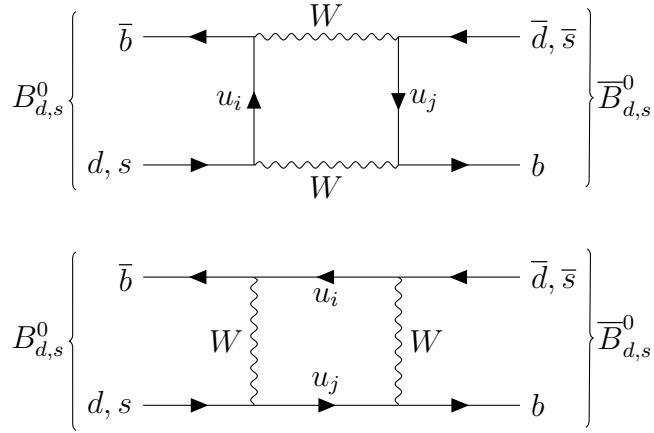
\begin{figure}[h]
		\centering
		\begin{tikzpicture}
			\begin{feynman}
				\vertex (a1) {\(\hspace{4.2mm} \overline b \)};
				\vertex[right=1.9cm of a1] (a2);
				\vertex[right=1.9cm of a2] (a3);
				\vertex[right=1.6cm of a3] (a4) {\(\overline d,\overline s\)};
				
				\vertex[below=1.7cm of a1] (b1) {\(d,s\)};
				\vertex[right=1.9cm of b1] (b2);
				\vertex[right=1.9cm of b2] (b3);
				\vertex[right=1.6cm of b3] (b4) {\(b\hspace{4mm}\)};
				
				\diagram* {
					{[edges=fermion]
						(b1) -- (b2) -- [edge label=\(u_i\)] (a2) -- (a1),
						(a4) -- (a3) --[edge label=\(u_j\)] (b3) -- (b4) ,
					},
					(a2) -- [boson, edge label=\(W\)] (a3),
					(b2) -- [boson, edge label'=\(W\)] (b3)
				};
				
				\draw [decoration={brace}, decorate] (b1.south west) -- (a1.north west)
				node [pos=0.5, left] {\(B_{d,s}^0\)};
				
				\draw [decoration={brace}, decorate] (a4.north east) -- (b4.south east)
				node [pos=0.5, right] {\(\overline{B}_{d,s}^0\)};
			\end{feynman}
		\end{tikzpicture}
		
		\vspace{5mm}
		
		\begin{tikzpicture}
			\begin{feynman}
				\vertex (a1) {\(\hspace{4.2mm} \overline b\)};
				\vertex[right=1.9cm of a1] (a2);
				\vertex[right=1.9cm of a2] (a3);
				\vertex[right=1.6cm of a3] (a4) {\(\overline d,\overline s\)};
				
				\vertex[below=1.7cm of a1] (b1) {\(d,s\)};
				\vertex[right=1.9cm of b1] (b2);
				\vertex[right=1.9cm of b2] (b3);
				\vertex[right=1.6cm of b3] (b4) {\(b\hspace{4 mm}\)};
				
				\diagram* {
					{[edges=fermion]
						(b1) -- (b2) --[edge label=\(u_j\)] (b3) -- (b4),
						(a4) -- (a3) --[edge label=\(u_i\)] (a2) -- (a1) ,
					},
					(b2) -- [boson, edge label=\(W\)] (a2),
					(b3) -- [boson, edge label'=\(W\)] (a3)
				};
				
				\draw [decoration={brace}, decorate] (b1.south west) -- (a1.north west)
				node [pos=0.5, left] {\(B^0_{d,s}\)};
				
				\draw [decoration={brace}, decorate] (a4.north east) -- (b4.south east)
				node [pos=0.5, right] {\(\overline{B}^0_{d,s}\)};
			\end{feynman}
		\end{tikzpicture}
		\caption{As figure \ref{figure1}, the leading contributions to $B^0_{d,s}-%
			\overline{B}^0_{d,s}$.}
		\label{figure2}
	\end{figure}

	
	For $K^0-\overline{K}^0$ and $B_{d,s}^0-\overline{B}_{d,s}^0$
	mixings, and given the fact that the valence quarks of these neutral mesons
	are all down-type, there will be no NP tree-level contributions to their
	mixing. Nonetheless, there are loop-level diagrams which may compete with
	the SM contributions. These box diagrams are presented in figures \ref
	{figure1} and \ref{figure2}. The off-diagonal component of the dispersive
	part of their amplitudes can be written as \cite{Cacciapaglia_2012}
	
	\begin{equation}
		(M_{12}^{N})^{*}\simeq \frac{m_{N}}{3\sqrt{2}}G_{F} f_{N}^{2} B_{N}\frac{\alpha }{4\pi s_{W}^{2}}\sum_{i,j=c,t,T}\eta _{ij}^{N} \lambda
		_{i}^{N}\lambda _{j}^{N}\ S(x_{i},x_{j}),  \label{m12}
	\end{equation}
	
	with the values of the bag parameters $B_{N}$, the decay constants $f_{N}$
	and the average masses $m_{N}$ for each meson presented in table \ref
	{table:1} and $G_{F}$ being the Fermi constant. Then, for the $B_{d,s}^0$
	system, the mass differences can be approximated as $\Delta m_{N}\simeq
	2|M_{12}^{N}|$, where the SM contributions are given by \cite{Branco:1999fs}
	
	\begin{equation}
	        \Delta m_{N}^{\text{SM}}\simeq \frac{G_{F}^{2} M_{W}^{2} m_{N}  f_{N}^{2}
			B_{N}}{6\pi ^{2}}  |\eta _{cc}^{N} S_{c} (\lambda _{c}^{N})^{2}+2\eta
		_{ct}^{N} S_{ct} \lambda _{c}^{N}\lambda _{t}^{N}+\eta _{tt}^{N} S_{t}
		(\lambda _{t}^{N})^{2}|.
	  \label{mnsm}
	\end{equation}
	
	The NP contribution is given by
	
	\begin{equation}
	        \Delta m_{N}^{\text{NP}}\simeq \frac{G_{F}^{2} M_{W}^{2} m_{N} f_{N}^{2}
			B_{N}}{6\pi ^{2}}  |2\eta _{cT}^{N} S_{cT} \lambda _{c}^{N}\lambda
		_{T}^{N}+2\eta _{tT}^{N} S_{tT} \lambda _{t}^{N}\lambda _{T}^{N}+\eta
		_{TT}^{N} S_{T} (\lambda _{T}^{N})^{2}|.
		  \label{mnnp}
	\end{equation}
	
	In Eqs. (\ref{m12}-\ref{mnnp}) we have defined
	
	\begin{equation}
	    \begin{array}{ccccc}
	        \lambda _{i}^{K}\equiv V_{is}^{*}V_{id}, & &
	        \lambda
		_{i}^{B_{d}}\equiv V_{ib}^{*}V_{id}, & &
		\lambda _{i}^{B_{s}}\equiv
		V_{ib}^{*}V_{is},
	    \end{array}
    \label{la}
	\end{equation}
	and introduced the Inami-Lim functions \cite{Inami:1980fz} $S_{ij}\equiv
	S(x_{i},x_{j})$ and $S_{i}\equiv S(x_{i})$ with $x_{i}\equiv
	(m_{i}/m_{W})^{2}$. The explicit
	expressions for these functions are presented in Appendix \ref{appendix_B}. We also use the approximation $x_{u}\simeq 0$ and the conditions
	
	\begin{equation}
	    \lambda^N_u+\lambda^N_c+\lambda^N_t+\lambda^N_T=0,
	\end{equation}
	which arise from the unitarity of the columns of $V^{CKM}$, allowing one, from this expression, to substitute the up-quark contributions.
	
    The masses $%
	m_{i}$ which enter these expressions are the $\overline{\text{MS}}$ masses $%
	m_{i}(\mu =m_{i})$. For the SM quarks in these processes, we use the central
	values \cite{Chetyrkin_2009}, \cite{Huang_2020} of 
	
	\begin{equation}
	    \begin{array}{c}
	        m_{c}(m_{c})=1.279\pm 0.013\text{ GeV},  \\
	        \\
	        m_{t}(m_{t})=162.6\pm 0.4
		\text{ GeV}. 
	    \end{array}
	\end{equation}
	
	The factors $\eta^N_{ij}$ account for $\mathcal{O}(1)$ QCD corrections to
	these electroweak interactions. Henceforth, we use the central values
	presented in \cite{Buras_2010,Brod_2010,Bobeth_2017}
	
	\begin{equation}
		\begin{array}{c}
			\eta _{tt}^{K}=0.5765\pm 0.0065\hspace{0.5mm}, \\ 
			\\
			\eta _{ct}^{K}=0.496\pm
			0.04\hspace{0.5mm},  \\ 
			\\
			\eta _{tt}^{B}=0.55\pm 0.01.
		\end{array}
		\label{eta1}
	\end{equation}

	For the remaining correction factors associated with the $B_{d,s}^0$ systems
	we use $\eta_{ij}^{B}\simeq 1$, which should not be problematic, given that
	the terms in Eq. (\ref{mnsm}) and Eq. (\ref{mnnp}) to which they are
	associated, are not relevant in calculations. In fact, in these processes,
	the terms in $(\lambda _{t}^{N})^{2}$ will dominate the SM contribution,
	whereas the term in $\lambda _{t}^{N}\lambda _{T}^{N}$ will dominate the NP
	contribution. Following \cite{Buras_2010}, the QCD corrections involving $T$
	shall be approximated as
	
	\begin{equation}
		\begin{array}{ccccc}
			\eta _{cT}^{K}\simeq 
			\eta _{ct}^{K},& &
			\eta _{tT}^{K}\simeq
			\eta_{TT}^{K}\simeq
			\eta _{tt}^{K}, & &
			\eta _{tT}^{B}\simeq \eta_{TT}^{B}\simeq
			\eta_{tt}^{B}.
		\end{array}
		\label{eta2}
	\end{equation}
	
	Assuming that $s_{24}=s_{34}=0$, we now obtain for the ratio of the
	NP-contribution versus of the SM-contribution:
	
	
	\begin{equation}
		\delta m_{B_{i}}\equiv \frac{\Delta m_{B_{i}}^{\text{NP}}}{\Delta m_{B_{i}}^{%
				\text{SM}}}\simeq \left| \frac{2S_{tT}\ \lambda _{t}^{B_{i}}\lambda
			_{T}^{B_{i}}}{S_{t}(\lambda _{t}^{B_{i}})^{2}}\right| \simeq 2s_{14}^{2}%
		\frac{S_{tT}}{S_{t}}\frac{|V_{ui}||V_{ub}|}{|V_{ti}||V_{tb}|}  \label{dmB}
	\end{equation}
	with $i=d,s$ and $c_{14}\simeq 1$. Then, inserting in this expression a
	value for $s_{14}\simeq 0.04$ and the current best-fit values for the moduli
	of the CKM entries (for the case of non-unitarity \cite{Zyla:2020zbs}) one
	finds each $\delta m_{B_{i}}$ to be a very slowly growing function with $%
	m_{T}$, and even at extremely large masses, the NP contributions will be
	very suppressed. For instance at $m_{T}=10$ \text{TeV}, one has $\delta
	m_{B_{d}}\simeq 0.681\%$ and $\delta m_{B_{s}}\simeq 0.032\%$. Hence, our
	model is safe with regard to both $\Delta m_{B_{d}}$ and $\Delta m_{B_{s}}$.
	
	$\Delta m_{K}^{\text{SM}}$ is long-distance dominated and up to now, still,
	there is no definite calculation of this quantity. Nevertheless the NP
	contribution is short-distance dominated and we can use Eq. (\ref{mnnp}). A
	reasonable constrain is therefore
	
	\begin{equation}
		\Delta m_{K}^{\text{NP}}<\Delta m_{K}^{\text{exp}},
	\end{equation}
	which for $s_{14}\simeq 0.04$ implies that $m_{T}<3.2$ TeV $\sim 20m_{t}$.
	Thus, below this very large upper bound for $m_{T}$, we may consider the
	model safe with regard to $\Delta m_{K}$.
	
\begin{table*}
\centering
\label{table:1}       
\begin{tabular}{ccccc}
\hline\noalign{\smallskip}
$N$ & $m_N$ [MeV] & $\Delta m^\text{exp}_N$ [MeV] & $f_N$ [MeV] & $B_N$ \\
\noalign{\smallskip}\hline\noalign{\smallskip}
$K$ & $497.611\pm0.013$ & $(3.484 \pm 0.006)\times 10^{-12}$ & $%
			155.7\pm 0.3$ & $0.717 \pm 0.024$ \\ 
$B_d$ & $5279.65\pm0.12$ & $(3.334 \pm 0.013)\times 10^{-10}$ & $190.0\pm
			1.3$ & $1.30 \pm 0.10$ \\ 
$B_s$ & $5366.88\pm 0.14$ & $(1.1683 \pm 0.0013)\times 10^{-8}$ & $%
			230.3\pm 1.3$ & $1.35 \pm 0.06$ \\ 
\noalign{\smallskip}\hline
\end{tabular}
\caption{Mass and mixing parameters \protect\cite{Zyla:2020zbs} and decay constants and bag parameters \protect\cite{Aoki_2020} for the neutral meson systems.}
\end{table*}
	
\subsection{New Insights on $\epsilon _{K}$ in the decay $K_{L}\rightarrow
		\pi \pi $ and New Physics \label{eK}}
	
	In this subsection, we focus on the parameter $\epsilon _{K}$, which
	describes indirect CP violation in the neutral kaon system. We propose a
	more retrictive upper-bound on the contributions to $\epsilon _{K}$ from New
	Physics. This upper-bound poses serious constraints on New Physics models.
	
	This parameter is associated \cite{Buras_1998} with $M_{12}^{K}$ through
	\begin{equation}
		|\epsilon _{K}|=\frac{\kappa _{\epsilon }}{\sqrt{2}\Delta m_{K}}\ |\text{Im} \
		M_{12}^{K}|,  \label{ek}
	\end{equation}
	with $\kappa _{\epsilon }\simeq 0.92\pm 0.02$ \cite{Buras_2008}.
	
	The NP contribution is essentially given by
	\begin{equation}
	    |\epsilon^{\text{NP}}_{K}|\simeq \frac{G_{F}^{2} M_{W}^{2} m_{K}
			f_{K}^{2} B_{K} \kappa _{\epsilon }}{12\sqrt{2}\pi ^{2}\Delta m_{K}} \left| 
		\text{Im}\left[ 2\eta _{cT}^{K} S_{cT} \lambda _{c}^{K}\lambda
		_{T}^{K}+2 \eta _{tT}^{K} S_{tT} \lambda _{t}^{K}\lambda _{T}^{K}+\eta
		_{TT}^{K} S_{T} (\lambda _{T}^{K})^{2}\right] \right| ,  \label{ek1}
	\end{equation}
	which is a valid expression for parametrizations with real $\lambda _{u}^{K} 
	$, as in our BC parametrization. In the sequel, when computing the
	quantities in Eq. (\ref{ek1}) numerically, we use the experimental value of $%
	\Delta m_{K}$ in table \ref{table:1}.
	
	From Eqs. (\ref{vs14}, \ref{ek1}), one can easily obtain the exact
	expression for the exact $s_{14}$ dominance case:
	\begin{equation}
		|\epsilon^{\text{NP}}_{K}| =\frac{G_{F}^{2} M_{W}^{2} m_{K} f_{K}^{2}
			B_{K} \kappa _{\epsilon }}{12\sqrt{2}\pi ^{2}\Delta m_{K}} \mathcal{F},
		\label{eK2a}
	\end{equation}
	with
	\begin{equation}
		\mathcal{F}=(\eta _{tT}^{K}\ S_{tT}-\eta _{cT}^{K}\ S_{cT})\
		c_{12}c_{13}^{2}c_{23}s_{12}s_{13}s_{23}s_{14}^{2}\sin \delta .
	\end{equation}
	
	\subsection*{A new upper-bound for $|\epsilon^{\text{NP}}_{K}|$}
	
	At this point, we introduce a new upper-bound for $|\epsilon^{\text{NP}}_{K}|$, which is far more restrictive than one used until recently 
	\begin{equation}
		|\epsilon^\text{NP}_K|<|\epsilon^\text{exp}_K|.
	\end{equation}
	In a recent paper by Brod, Gorbahn and Stamou (BGS) \cite{Brod_2020} it was
	shown that through manifest CKM unitarity it was possible to circumvent the
	large uncertainties related to the charm-quark contribution to $\epsilon
	_{K} $, allowing for an SM prediction of $|\epsilon _{K}|$,
	\begin{equation}
		|\epsilon ^{\text{SM}}_{K}|=(2.16\pm 0.18)\times 10^{-3},  \label{eksm}
	\end{equation}
	which is very compatible with the experimental value $|\epsilon^{\text{%
			exp}} _{K}|$ $=(2.228\pm 0.011)\times 10^{-3}$, with a relative error of the order of $10\%$.  Thus, 
	\begin{equation}
	 |\epsilon^{\text{exp}}_{K}|-|\epsilon^{\text{SM}}_{K}|\simeq \left(0.68\pm 1.80\right)\times 10^{-4},
		\label{delta}
	\end{equation}
which we will use in the global analysis of section \ref{global_analysis}. 

At $1\sigma$ one may establish a new upper-bound for the NP contribution to $|\epsilon _{K}|$ such that $%
	|\epsilon^{\text{NP}}_{K}|\lesssim 0.1|\epsilon^{\text{exp}}_{K}|$, or more concretely
	\begin{equation}
		|\epsilon^{\text{NP}}_{K}|^{1\sigma}\ \lesssim \Delta= 2.48\times 10^{-4},
		\label{delta}
	\end{equation}
	which severely restricts various models, including the
	present one with exact $s_{14}$ dominance.
	
\begin{figure*}
\centering
\resizebox{0.85\textwidth}{!}{%
  \includegraphics[scale=0.65]{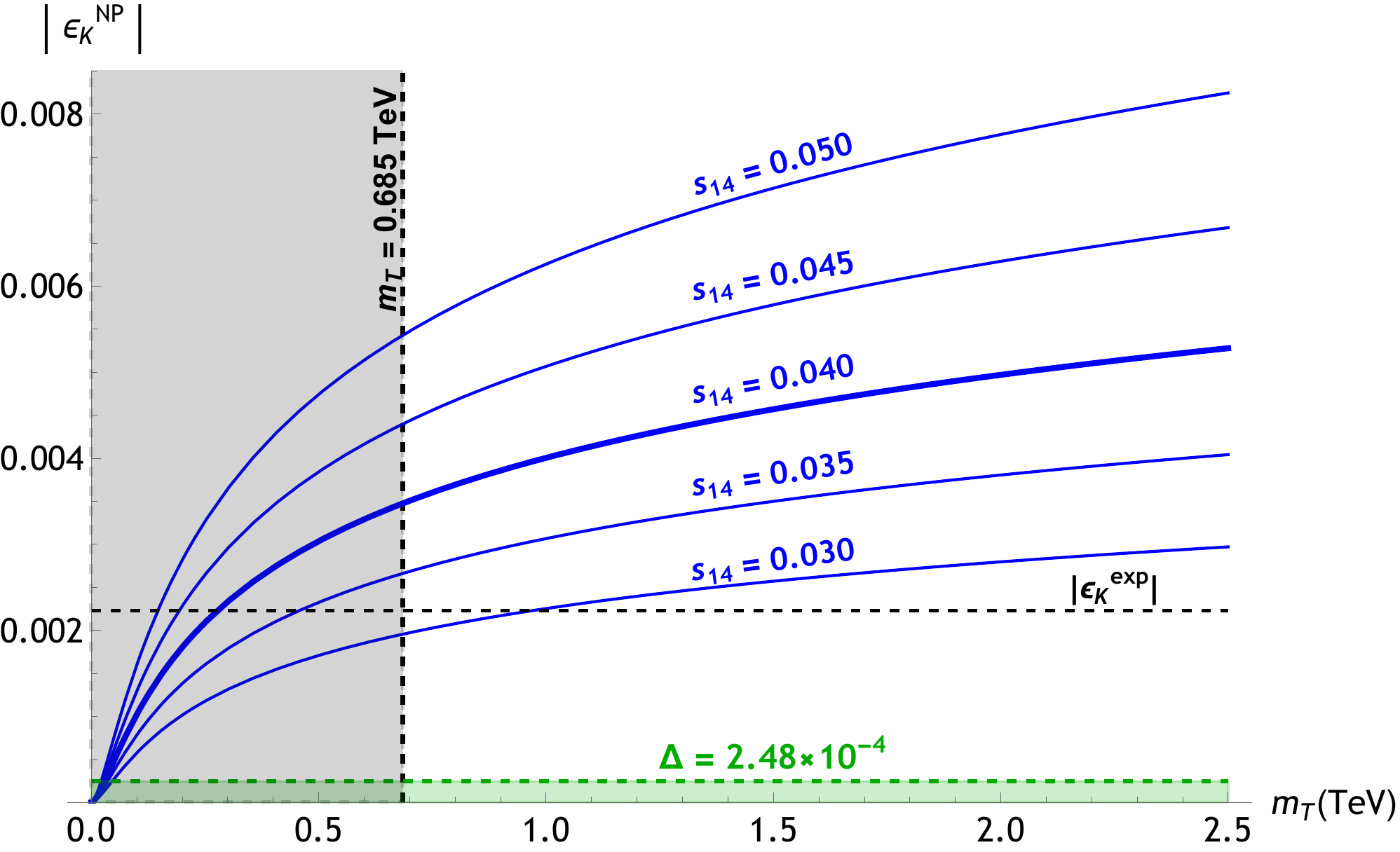}
}
\caption{$\vert\epsilon_K^\text{NP}\vert$ as a function of $m_T$ in the framework of strict $s_{14}$ dominance ($s_{24}=s_{34}=0$), for various values of $s_{14}$. The vertical line represents the experimental lower bound for the mass of the heavy top, $m_T>0.685$ TeV. The black horizontal line corresponds to $\vert\epsilon^\text{NP}_K\vert=\vert\epsilon^\text{exp}_K\vert$, whereas the green one corresponds to $\vert\epsilon^\text{NP}_K\vert=\Delta$. In green we represent the region inside the range of interest	for $m_T$ where the model might be safe.}	\label{figure3}       
\end{figure*}

	Using this, in figure \ref{figure3} we present a plot of Eq. (\ref{eK2a}) as a function
	of $m_{T}$ for various values of $s_{14}$ and 
	\begin{equation}
		\begin{array}{ccc}
		    \theta _{12}\simeq 0.2264, & &
			\theta _{13}\simeq 0.0037,\\
			\\
			\theta _{23}\simeq 0.0405, & &
			\delta \simeq 1.215.
		\end{array}
		\label{param}
	\end{equation}
	Note that only when $s_{14}\lesssim 0.03$ is one able to obtain $|\epsilon^{\text{NP}}
	_{K}|$ $\lesssim |\epsilon^{\text{exp}}_{K}|$. For larger values of $s_{14}$,
	one has mostly that $|\epsilon^{\text{NP}}_{K}|>|\epsilon^{\text{exp}} _{K}|$%
	. We conclude that our $1\sigma$ upper-bound on $%
	|\epsilon^{\text{NP}}_{K}|$ in Eq. (\ref{delta}) is only achieved in experimentally ruled out
	regions for $m_{T}$ and is incompatible with $s_{14}\simeq 0.04$. Thus, we
	find that the parameter region of exact $s_{14}$ dominance, where we
	strictly have that $s_{24}=s_{34}=0$, is not safe with regard to $|\epsilon
	_{K}|$. 

	However, in the next section, we will show that a small $|\epsilon^{%
		\text{NP}}_{K}|$ obeying $|\epsilon^{\text{NP}}_{K}|\ \lesssim \Delta $, is
	achievable, if the strict $s_{24}=s_{34}=0$ imposition is dropped and
	replaced by a more realistic one, where $s_{24},s_{34}\neq 0$, but with $%
	s_{24},s_{34}\ll s_{14}$. This slightly different framework, however, shares
	the same relevant features as the exact $s_{14}$ dominance case, without
	changing the pattern of decays and predictions for the heavy top.
	
	\subsection{Heavy $T-$ decays}
	
	As long as we have that, from all three extra angles, only the angle $s_{14}$
	differs from zero, the new heavy $T$ quarks get mixed with the $u$ quark. In
	the neutral currents, we have $\left| F_{14}^{u}\right| \sim s_{14}$
	controlling the decays $T\longrightarrow u\ Z$ and $T\longrightarrow u\ h$.
	In the charged currents, we have $\left| V_{Td}\right| \sim s_{14}$, $\left|
	V_{Ts}\right| \sim s_{14}\lambda $ and $\left| V_{Tb}\right| \sim
	s_{14}\lambda ^{3}$, from which one concludes that the dominant decay
	channel is $T\longrightarrow d\ W$. For the range of masses we consider, one
	has, to a very good approximation \cite{Botella_2017} 
	$$
	\Gamma \left( T\longrightarrow d\ W\right) \simeq 2\Gamma \left(
	T\longrightarrow u\ Z\right) \simeq 2\Gamma \left( T\longrightarrow u\
	h\right).
	$$
	For experimental purposes, these three decay channels to the light quarks
	dominate the total decay width. This dominance to light quark channels is a
	distinctive feature of the $s_{14}$ dominance scenario and is the origin of
	the fact that we can consider masses as light as $m_{T}$ $=685\ GeV$ \cite
	{Sirunyan_2018}. Note that major experimental searches correspond to the
	channels $\Gamma \left( T\longrightarrow b\ W\right) $, $\Gamma \left(
	T\longrightarrow t\ Z\right) $, $\Gamma \left( T\longrightarrow t\ h\right) $%
	, here highly suppressed.
	
	\section{Solving the $\epsilon _{K}$ problem while maintaining the main
		features of the $s_{14}-$ dominance case}
	
	As stated above, the strict imposition of $s_{24}=s_{34}=0$ above might be
	considered somewhat unnatural. A possible more realistic scenario would be
	one, with small, but non-zero $s_{24}$ and $s_{34}$. In this section, we
	give an analysis of the previous electroweak-precision-measurements (EWPM) related quantities allowing for small values $%
	s_{24},s_{34}\ll s_{14}$
	\begin{equation}
		s_{34} , s_{24} \lesssim \lambda ^{5}
		\label{s24s34}
	\end{equation}
	while still keeping our solution for CKM unitarity problem with $%
	s_{14}\simeq 0.04$. We show that it is possible to find a suitable solution
	for the $|\epsilon _{K}|$ problem described in the previous section \ref{eK}%
	, while preserving all the important features of the model, i.e. without
	significantly affecting predictions for other observables. In addition, we
	also point out that other important CP-violation quantities, in particular $%
	\epsilon ^{\prime }/\epsilon $ and $Br\left( K_{L}\longrightarrow \pi
	^0\nu \overline{\nu }\right) $, require new attention.
	
	\subsection{Modifications to the NP contributions in neutral meson mixings}
	
	Using the Botella-Chau parametrization, with $c_{13},c_{23},c_{24},$ $c_{34}
	\simeq 1$ and rephasing the left-handed heavy top quark field as $%
	T_{L}\rightarrow e^{i\delta _{14}}T_{L}$, we parametrize the CKM matrix, in
	leading order, as presented in Eq. (\ref{vckm2})
	where the $V_{ij}$ represent the $(i,j)$ entries of $\mathcal{V}^{CKM}$ in Eq. (\ref
	{vs14}). Here, we relax one of the upper-bounds in Eq. (\ref{s24s34}) and
	assume even that $\left| s_{34}\right| \lesssim \lambda ^{5}$ while $\left|
	s_{24}\right| \lesssim \lambda ^{4}$. We have also defined the difference $%
	\delta ^{\prime }\equiv \delta _{24}-\delta _{14}$ of the extra phases,
	which play a role futher on.
	
	    \begin{equation}
	    \begin{array}{l}
			\mathcal{V}^{CKM}
			=\\
			\\
			\small{\left( 
			\begin{array}{lll}
				V_{11} & V_{12} & V_{13} \\ 
				V_{21}-c_{12}s_{14}s_{24}e^{-i\delta ^{\prime }} & 
				V_{22}-s_{12}s_{14}s_{24}e^{-i\delta ^{\prime }} & V_{23} \\ 
				V_{31}-c_{12}s_{14}s_{34}e^{i\delta _{14}} & V_{32} & V_{33} \\ 
				V_{41}+s_{12}s_{24}e^{i\delta ^{\prime }} & V_{42}-c_{12}s_{24}e^{i\delta
					^{\prime }}-c_{12}s_{23}s_{34}e^{-i\delta _{14}} & V_{43}-s_{23}s_{24}e^{i%
					\delta ^{\prime }}-s_{34}e^{-i\delta _{14}}
			\end{array}
			\right)}\\
			\\+  \mathcal{O}(\lambda ^{8}),
	    \end{array}
		\label{vckm2}
	\end{equation}
	
	Instead of the expression given in Eq. (\ref{eK2a}), the overall NP
	contribution to $|\epsilon _{K}|$ is now approximated by
	
	\begin{equation}
		|\epsilon^{\text{NP}}_{K}| \simeq \frac{G_{F}^{2} M_{W}^{2} m_{K}
			f_{K}^{2} B_{K} \kappa _{\epsilon }}{12\sqrt{2}\pi ^{2}\Delta m_{K}}
		\left| \mathcal{F-F}^{\prime }\right| =q_{K} \left| \mathcal{F-F}^{\prime
		}\right| ,  \label{eK2}
	\end{equation}
	with $\mathcal{F}^{\prime }$, being an extra contribution to $|\epsilon^{\text{NP}}
	_{K}|$ coming from the fact that $s_{24},s_{34}\neq 0$.
	
	It is worthwhile to give an approximate expression for this new $|\epsilon
	_{K}|^{\text{NP}}$, in terms of our BC parametrization in Eqs. (\ref{vs14}, 
	\ref{vckm2}). In leading order, one finds for Eq. (\ref{eK2}),
	
	\begin{equation}
	        |\epsilon^{\text{NP}}_{K}|=  2 q_{K} s_{12} s_{14}^{2} \left| \eta
		_{tT}^{K} S_{tT} s_{13}s_{23}\sin \delta -\eta _{TT}^{K} S_{TT}
		s_{14}s_{24}\sin \delta ^{\prime }\right|.
		\label{F2}
	\end{equation}
	
	Note that this leading order contribution to $|\epsilon^{\text{NP}}_{K}|$
	is only dependent on the phase combination $\delta ^{\prime }=\delta
	_{24}-\delta _{14}$ and is independent of $s_{34}$, because we chose $%
	s_{34}\leq \lambda ^{5}$. In fact, this also true for the next-leading order
	terms.
	
	From Eq. (\ref{F2}), it is already clear that $|\epsilon^{\text{NP}}_{K}|$
	may become small in certain regions of parameter-space, if the two terms in
	the expression can cancel each other. Moreover, if we restrict ourselves to
	a region of the mass $m_{T}$ (of the extra heavy quark) between $5m_{t}\leq
	m_{T}\leq 12m_{t} $, then with Eqs. (\ref{eta1}, \ref{eta2}), we find that $%
	\eta _{tT}^{K}\ S_{tT}$ and $\eta _{TT}^{K}\ S_{TT}$ in Eq. (\ref{F2})
	behave, in a good approximation, as linear functions of $k=\frac{m_{T}}{m_{t}%
	}$%
	\begin{equation}
		\begin{array}{c}
		    \eta _{tT}^{K} S_{tT}\approx 2.492+0.1492 \ k, \\
		    \\
			\eta _{TT}^{K}
			S_{TT}\approx -36.613+10.232 \ k.
		\end{array}
		\label{etA}
	\end{equation}
	With this simplification and with the PDG values for $s_{13}$, $s_{23}$ as
	well as our proposed value for $s_{14}\approx 0.04$, one finds that there exists a
	fairly large parameter region (depending on $\theta _{24}$) which is allowed
	for $\delta ^{\prime }$ and where $\delta ^{\prime }\in \left[
	1.0,2.0\right] $. Thus, we find that this new phase $\delta ^{\prime }$
	assumes values, in this context, which are very similar to the usual
	CP-violating phase $\delta $.
	
\begin{figure*}
\centering
\resizebox{0.85\textwidth}{!}{%
  \includegraphics{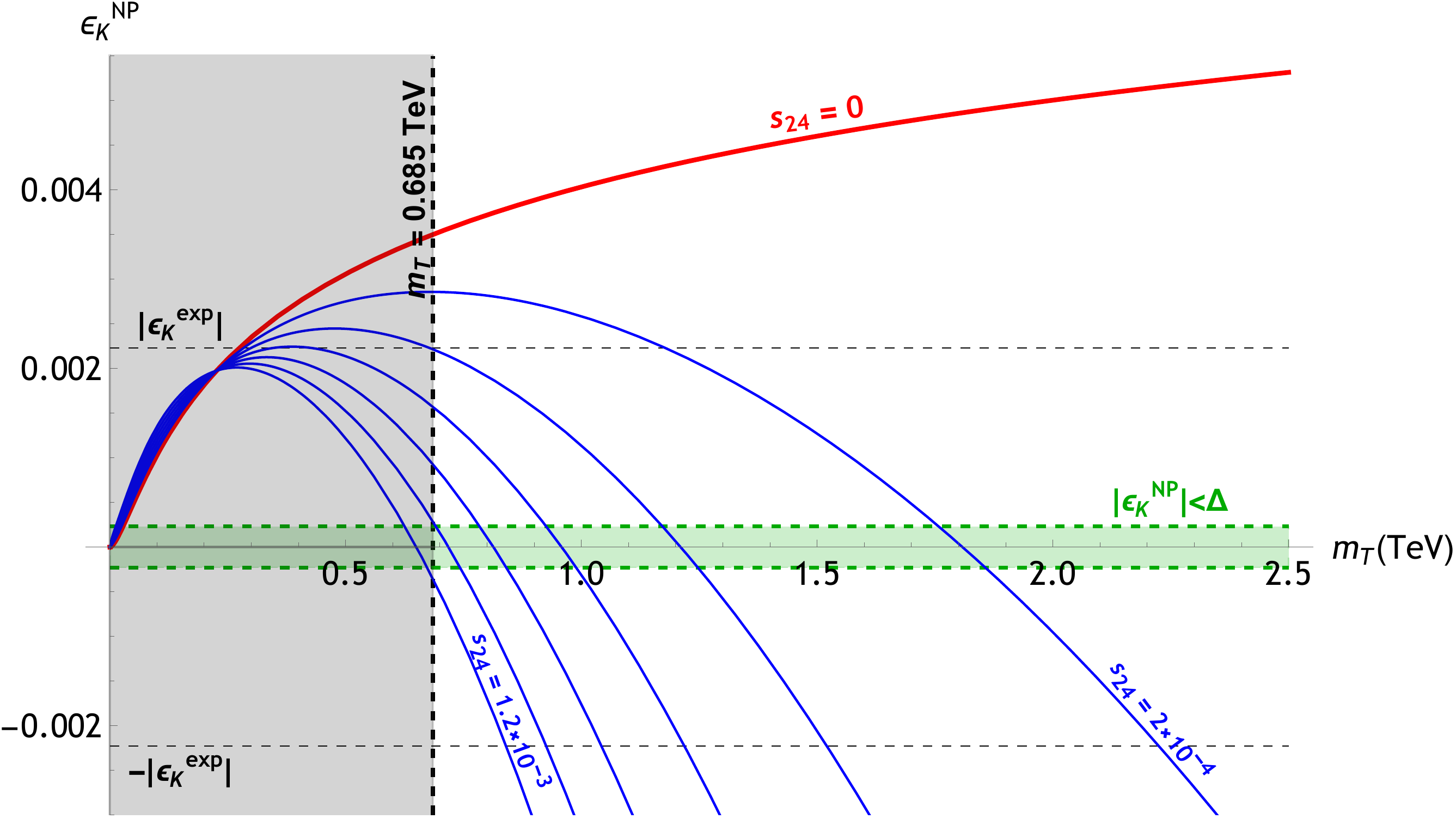}
}
\caption{Analogous plot to that of figure \ref{figure3} but for realistic dominance with $\theta_{14}=0.04$ and $\delta^{\prime}= \pi/2$. Various	values of $s_{24}$ are spanned in steps of $2\times 10^{-4}$ for $s_{24}\in[0,1.2\times 10^{-3}]$. The curve for $s_{24}=0$ (in red) is the thicker one in figure \ref{figure3}. The region $|\epsilon_K|^\text{NP}<\Delta$ is highlighted in green for $m_T>0.685$ TeV.}
		\label{plot2}       
\end{figure*}

	In figure \ref{plot2}, we plot Eq. (\ref{eK2}) for various values of $s_{24}$%
	, using Eq. (\ref{param}) with $s_{14}=0.04$ and a central value for $\sin
	\delta ^{\prime }=1$. From the plot we conclude that small values of $%
	|\epsilon^{\text{NP}}_{K}|$ can be achieved, e.g. for $m_{T}=0.685$ TeV by
	having $s_{24}\lesssim 1.2\times 10^{-3}$, or e.g. for $m_{T}=1.0$ TeV by
	having $s_{24}\lesssim 6\times 10^{-4}$. Thus, we find a region where the problem
	discussed in section $\ref{eK}$ can be fixed. In addition, one can see that
	having $s_{24}<2\times 10^{-4}$ is undesirable as it would require very
	large heavy top masses ($m_{T}\gtrsim 2$ TeV) to achieve $|\epsilon _{K}|^{%
		\text{NP}}\lesssim \Delta .$
	
	The NP contributions to $\Delta m_{B_{i}}$ will also be modified, with all
	changes coming essentially from $\lambda _{T}^{B_{i}}$. From Eq. (\ref{vckm2}%
	) one finds for $s_{34}\sim \lambda ^{5}$, in leading order
	\begin{equation}
	\begin{split}
	     \lambda _{T}^{B_{d}}\simeq V_{41}V_{43}^{*}\left( 1-\frac{s_{34}}{V_{43}^{*}}%
		e^{i\delta _{14}}\right) ,  \\
		\\
	     \lambda _{T}^{B_{s}}\simeq
		V_{42}V_{43}^{*}\left( 1-\frac{s_{34}}{V_{43}^{*}}e^{i\delta _{14}}\right) , 
	\end{split}
		\label{lambBT}
	\end{equation}
	so that now, we have an extra term for each quantity which competes with the
	absolute dominance result. For $s_{34}\sim |V_{43}|\sim \lambda ^{5}$ the new
	term will be of the order of the old one, which should not be problematic
	given how insignificant the NP contributions to $\Delta m_{B_{i}}$ are in
	the absolute dominance framework.
	
	Still, if one requires that in this alternative framework the predictions
	for $\Delta m^\text{NP}_{B_{i}}$ do not differ significantly from the ones of absolute
	dominance, then Eq. (\ref{lambBT}) seems to favor $s_{34}\ll |V_{43}|\sim
	\lambda ^{5}$ and we are able to recover the results of absolute dominance.
	This fact,
	when coupled with the independence of Eq. (\ref{F2}) on $s_{34}$ suggests
	that the $s_{14}$ dominance framework might be viable even with $s_{34}\ll s_{24}$. On the
	other hand, the observable $\Delta m_{K}$ will not be meaningfully altered
	when switching to Eq. (\ref{s24s34}) as the new terms in Eq. (\ref{vckm2})
	which contribute to $\lambda _{T}^{K}$ are dominated by $|V_{41}|\sim \lambda
	^{2}$ and $|V_{42}|\sim \lambda ^{3}$. 
	
	\subsection{Emergence of more New Physics \label{NNP}}
	
	Having non-zero $s_{24}$ and $s_{34}$ implies non-zero $\mathcal{V}_{42}$
	and $\mathcal{V}_{43}$ which in turn will induce NP contributions to $D^0-%
	\overline{D}^0$ mixing and allow rare decays of the top quark into the
	lighter generations, which was not true before. We now will briefly study
	these processes, as well as others\footnote[1]{For more possible effects see also \cite{Balaji:2021lpr}.}, like $K_{L}\rightarrow \pi ^0\overline{%
		\nu }\nu $ and the CP violation observable $\epsilon ^{\prime }/\epsilon $.
	
	\subsection*{$D^0-\overline{D}^0$ mixing}
	
	The NP tree-level contribution to the $D^0-\overline{D}^0$ mixing is
	described by the effective Lagrangian \cite{Branco_1995}
	
	\begin{equation}
		\mathcal{L}_{\text{eff}}^{\text{NP}}=-\frac{G_{F}}{\sqrt{2}}\ (\mathcal{V}%
		_{41}^{u*}\mathcal{V}_{42}^{u})^{2}\ (\overline{u}_{L}\gamma ^{\mu }c_{L})(%
		\overline{u}_{L}\gamma _{\mu }c_{L}),  \label{ff}
	\end{equation}
	This results in a contribution to the $D^0$ mixing parameter $x_{D}\equiv
	\Delta m_{D}/\Gamma _{D}$ given by \cite{Golowich_2009}
	
	\begin{equation}
		x_{D}^{\text{NP}}\simeq \frac{\sqrt{2}m_{D}}{3\Gamma _{D}}\ G_{F}\
		f_{D}^{2}\ B_{D}\ r(m_{c},M_{Z})\ |\mathcal{V}_{41}^{u*}\mathcal{V}%
		_{42}^{u}|^{2}.  \label{gg}
	\end{equation}
	where $r(m_{c},M_{Z})\simeq 0.778$ is a factor that accounts for RG effects.
	The remaining constants are $m_{D}=1864.83\pm 0.05$ MeV, $\Gamma _{D}=1/\tau
	_{D}$ with $\tau _{D}=(410.1\pm 1.5)\times 10^{-15}$ s \cite{Zyla:2020zbs}, $%
	B_{D}=1.18_{-0.05}^{+0.07}$ \cite{Buras_2010_2} and $f_{D}=212.0\pm 0.7$ MeV 
	\cite{Aoki_2020}. Requiring $s_{24}\lesssim \lambda ^{5}$ yields an upper
	bound for the NP contribution of $x_{D}^{\text{NP}}<0.015\%$, which is
	negligible when compared to the experimental value, $x_{D}^{\text{exp}%
	}=0.39_{-0.12}^{+0.11}\%$ \cite{Amhis_2021}.
	
	\begin{figure}[h]
		\centering
		\begin{tikzpicture}
			\begin{feynman}
				\vertex (a1) {\(\overline u\)};
				\vertex[right=1.5cm of a1] (a2);
				\vertex[right=2cm of a2] (a3);
				\vertex[right=1.2cm of a3] (a4) {\(\overline c \hspace{0.6mm}\)};
				
				\vertex[below=1cm of a2] (c1);
				\vertex[below=1cm of a3] (c2);
				
				\vertex[below=2cm of a1] (b1) {\(\hspace{0.6mm}c\)};
				\vertex[right=1.5cm of b1] (b2);
				\vertex[right=2cm of b2] (b3);
				\vertex[right=1.2cm of b3] (b4) {\(u\)};
				
				\diagram* {
					{[edges=fermion]
						(b1) -- (c1), 
						(c2) -- (b4),
						(a4) -- (c2),
						(c1) -- (a1) ,
					},
					(c1) -- [boson, edge label=\(Z\)] (c2),
				};
				
				\draw [decoration={brace}, decorate] (b1.south west) -- (a1.north west)
				node [pos=0.5, left] {\(D^0\)};
				
				\draw [decoration={brace}, decorate] (a4.north east) -- (b4.south east)
				node [pos=0.5, right] {\(\overline{D}^0\)};
			\end{feynman}
		\end{tikzpicture}
		\caption{NP contribution to $D^0 -\overline{D}^0$ mixing via $Z$-mediated
			FCNC.}
		\label{figure5}
	\end{figure}
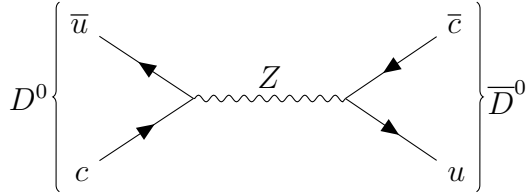
	
	\subsection*{Rare $t\rightarrow qZ$ decays}
	
	With $s_{34}\neq 0$, the mixing of the VLQ with the lighter generations will
	result in rates for the processes $t\rightarrow q_{i}Z$, ($q_{i}=u,c$) which
	may differ significantly from the ones predicted by the SM. In fact, the
	leading-order NP contribution occurs at tree-level and is given by \cite
	{Aguilar-Saavedra:2004mfd}
	\begin{equation}
	     	\Gamma (t\rightarrow q_{i}Z)_{\text{NP}}\simeq \frac{\alpha }{%
			32s_{W}^{2}c_{W}^{2}} |\mathcal{V}_{4i}^{u*}\mathcal{V}_{43}^{u}|^{2} \frac{m_{t}^{3}}{M_{Z}^{2}} \left( 1-\frac{M_{Z}^{2}}{m_{t}^{2}}\right)^{2}
		\left( 1+2\frac{M_{Z}^{2}}{m_{t}^{2}}\right).
	  \label{rr}
	\end{equation}
	
	Approximating the total decay width of the top-quark by $\Gamma_t\simeq%
	\Gamma(t\rightarrow bW^+)$, the branching ratio is 
	\begin{equation}
	        \text{Br}(t\rightarrow q_iZ)_\text{NP}\simeq \frac{|\mathcal{V}_{4i}^{u\ast}%
			\mathcal{V}_{43}^{u}|^{2}}{2|V_{33}|^{2}}\left( 1-\frac{M_{Z}^{2}}{m_{t}^{2}}%
		\right) ^{2} \left( 1+2\frac{M_{Z}^{2}}{m_{t}^{2}}\right)\left( 1-3\frac{%
			M_{Z}^{4}}{m_{t}^{4}}+2\frac{M_{Z}^{6}}{m_{t}^{6}}\right)^{-1}.
      \label{tt}
	\end{equation}
	
	However, for $s_{24}, s_{34}\lesssim \lambda^5$, it will never come close to
	exceed the experimental upper bounds: $\text{Br}(t\rightarrow uZ)_\text{exp}%
	< 1.7\times 10^{-4}$, $\text{Br}(t\rightarrow cZ)_\text{exp}< 2.4\times
	10^{-4}$ (95\% CL) \cite{2018}. For $s_{34}\sim 10^{-7}$, one might even
	conceivably achieve NP contributions lower than the SM predictions $\text{Br}%
	(t\rightarrow uZ)_\text{SM}\sim 10^{-16}$, $\text{Br}(t\rightarrow cZ)_\text{%
		SM}\sim 10^{-14}$ \cite{Aguilar-Saavedra:2004mfd}.
	
	\subsection*{The decay $K_{L}\rightarrow \pi^0\overline{\nu }\nu $}
	
	For this process, it is relevant to study the quantity $L$ proportional to
	the decay amplitude, which in the SM and using the standard PDG
	parametrization, can be written as \cite{Branco:1999fs}
	
	\begin{equation}
		L_{SM}=|\text{Im}\left[ \lambda _{c}^{K}\ X(x_{c})+\lambda _{t}^{K}\
		X(x_{t})\right] |^{2},  \label{aSM}
	\end{equation}
	where we have introduced an extra Inami-Lim function $X(x_{i})$, presented
	in Eq. (\ref{X_IL}) of Appendix B.
	
	When the heavy-top is introduced two new terms should be added to Eq. (\ref
	{aSM}), leading to
	
	\begin{equation}
		L=|\text{Im}\left[ \lambda _{c}^{K}\ X(x_{c})+\lambda _{t}^{K}\
		X(x_{t})+\lambda _{T}^{K}\ X(x_{T})+A_{ds}\right] |^{2}.  \label{aNP}
	\end{equation}
	
	The first term is a simple generalisation of the terms in Eq. (\ref{aSM})
	which is to be expected from the introduction of a new quark, whereas the
	last one accounts for the decoupling behaviour that arises from the fact
	that this new quark is an isosinglet and is responsible for generating
	FCNC's at tree level in the electroweak sector. Note that the
	gauge-invariant function in Eq. (\ref{X_IL}) is obtained by considering all
	diagrams that contribute to processes such as $K_{L}\rightarrow \pi ^0%
	\overline{\nu }\nu $, with some of these diagrams being $Z$-exchange penguin
	diagrams where we can have up-type quarks running inside a loop coupled to a 
	$Z$-boson, $i.e$ where the new FCNC's effects in the up quark sector have to
	be taken into account. The role of $A_{ds}$ is, therefore, to account for
	these effects.
	
	With regard to $A_{ds}$, we have \cite{Botella_2017_a}--\cite{Picek_2008}
	
	\begin{equation}
		A_{ds}=\sum_{i,j=c,t,T}V_{is}^{*}\left( F^{u}-I\right)
		_{ij}V_{jd}N(x_{i},x_{j}),
	\end{equation}
	with
	\begin{equation}
		\begin{split}
			N(x_{i},x_{j})=\frac{x_{i}x_{j}}{8}\left( \frac{\log x_{i}-\log x_{j}}{%
				x_{i}-x_{j}}\right) ,\\
            \\				
			N(x_{i},x_{i})\equiv \lim_{x_{j}\rightarrow	x_{i}}N(x_{i},x_{j})=\frac{x_{i}}{8}.
		\end{split}
	\end{equation}
	
	For $s_{24}\neq0$ and in the limit $s_{34}=0$, the FCNC-matrix $F^u$ in Eq. (%
	\ref{fcnc}) gets modified into
	\begin{equation}
	    \small
		F^{u}=\left( 
		\begin{array}{cccc}
			c_{14}^{2} & -c_{14}s_{14}s_{24}e^{i\delta ^{\prime }} & 0 & 
			-c_{14}c_{24}s_{14}e^{-i\delta _{14}} \\ 
			-c_{14}s_{14}s_{24}e^{-i\delta ^{\prime }} & 1-c_{14}^{2}s_{24}^{2} & 0 & 
			-c_{14}^{2}c_{24}s_{24}e^{-i\delta _{24}} \\ 
			0 & 0 & 1 & 0 \\ 
			-c_{14}c_{24}s_{14}e^{i\delta _{14}} & -c_{14}^{2}c_{24}s_{24}e^{i\delta
				_{24}} & 0 & 1-c_{14}^{2}c_{24}^{2}
		\end{array}
		\right) ,  \label{fcnca}
	\end{equation}
	So that to a very good approximation one can write
	\begin{equation}
		A_{ds}\simeq -\frac{ x_T}{8}c^2_{14}c^2_{24}\lambda^K_T.  \label{Ads}
	\end{equation}
	
	Thus, we obtain 
	\begin{equation}
		L\simeq |\text{Im}\left[ \lambda _{c}^{K}X(x_{c})+\lambda
		_{t}^{K}X(x_{t})+\lambda _{T}^{K}\Tilde{X}(x_T)\right] |^{2},  \label{aNPa}
	\end{equation}
	where, with $c_{14},c_{24}\simeq 1$, we have defined
	\begin{equation}
	        \Tilde{X}(x_T) \equiv X(x_{T})-\frac{x_{T}}{8}=\frac{x_{T}}{8(x_{T}-1)}\left(
		3+\frac{3x_{T}-6}{x_{T}-1}\log x_{T}\right) ,
		  \label{tildeX_IL}
	\end{equation}
	which shows the logarithmic behaviour of the NP piece\footnote[1]{%
		The piece linear in $x_T$ in $X(x_T)$ (see eq. (\ref{X_IL})) is not
		completely eliminated, but what survives the cancellation with $A_{ds}$ is
		suppressed by a factor of $s^2_{14}$, making it only relevant at very large
		masses.}. From Eq. (\ref{vs14}) it is clear that in the limit $s_{24}=0$
	there is essentially no NP piece, given that $\text{Im}\lambda _{T}^{K}=0$.
	However, if one takes $s_{24}\neq 0$ in order to fix the $\epsilon _{K}$
	problem, this is no longer true as $\text{Im}\lambda _{T}^{K}\simeq
	-c_{12}^{2}s_{14}s_{24}\sin \delta ^{\prime }$. In the considered range of
	parameters, we can get, in general, an important reduction of the branching
	ratio of the CP violation decay $K_{L}\rightarrow \pi ^0\overline{\nu }\nu 
	$
	\begin{equation}
		0.2\lesssim \frac{L}{L_{\text{SM}}}\simeq \frac{\text{Br}\left(
			K_{L}\rightarrow \pi ^0\overline{\nu }\nu \right) }{\text{Br}\left(
			K_{L}\rightarrow \pi ^0\overline{\nu }\nu \right) _{\text{SM}}}\lesssim 0.8.
	\end{equation}
	
	\subsection*{The decay $K^{+}\rightarrow \pi ^{+}\overline{\nu }\nu $}
	
	Similarly, this process is studied analysing the ratio
	\begin{equation}
		\frac{L^+}{L^+_\text{SM}}\equiv  \frac{\text{Br}(K^{+}\rightarrow \pi ^{+}\overline{\nu }\nu )}{\text{Br}%
			(K^{+}\rightarrow \pi ^{+}\overline{\nu }\nu )_{\text{SM}}}=\left| \frac{%
			\lambda _{c}^{K}X^{\text{NNL}}(x_{c})+\lambda _{t}^{K}X(x_{t})+\lambda
			_{T}^{K}X(x_{T})+A_{ds}}{\lambda _{c}^{K}X^{\text{NNL}}(x_{c})+\lambda
			_{t}^{K}X(x_{t})}\right| ^{2},
			\label{BrKp}
	\end{equation}
	where, here, the charm contribution cannot be overlooked, because, even
	though $X^{\text{NNL}}(x_{c})\ll X(x_{t})$, one has that $\lambda
	_{c}^{K}\gg \lambda _{t}^{K}$. Also, instead of the previous charm
	contribution $X(x_{c})$, we now use the NNLO \cite{Buras_2015} charm
	contribution $X^{\text{NNL}}(x_{c})\simeq 1.04\times 10^{-3}$ (see Appendix
	B).
	
\begin{figure*}
	\centering
\resizebox{0.9\textwidth}{!}{%
  \includegraphics{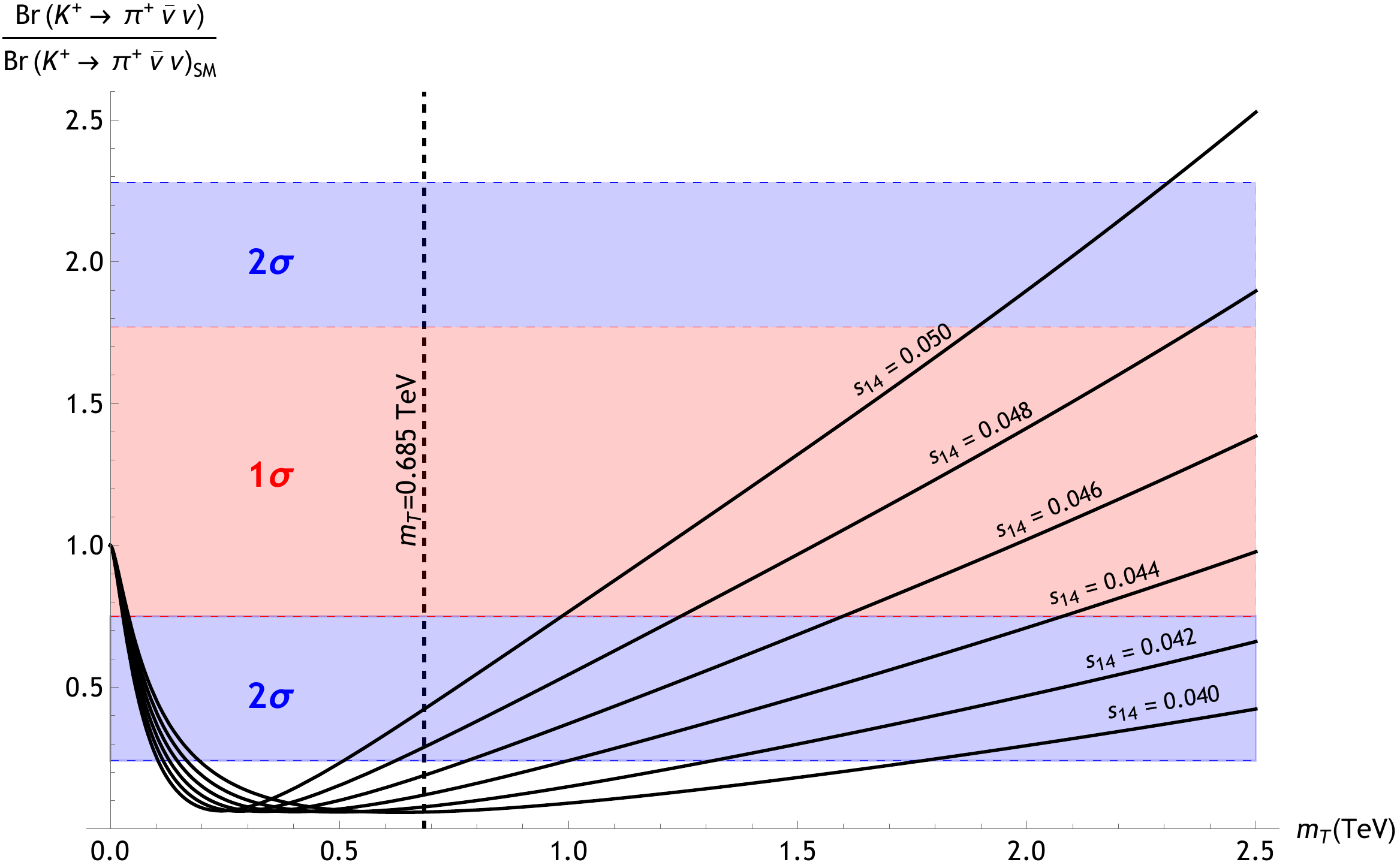}
}
\caption{Plot of Eq. (\ref{BrKp}) as a function of $m_T$ for various values of $s_{14}\geq 0.04$ with $s_{24}=s_{34}=5\times 10^{-4}$, $\delta^\prime=1.2 $ and using Eq. (\ref{param}). The coloured regions refer to the $1\sigma$ and $2\sigma$ ranges.}
	\label{BrKp_plot}     
\end{figure*}
	
	Current measurements of this decay yield $\text{Br}(K^{+}\rightarrow \pi ^{+}%
	\overline{\nu }\nu )_{\text{exp}}=\left( 10.6_{-3.4}^{+4.0}\pm 0.9\right)
	\times 10^{-11}$, whereas the SM prediction is $\text{Br}(K^{+}\rightarrow
	\pi ^{+}\overline{\nu }\nu )_{\text{SM}}=\left( 8.4\pm 1.0\right) \times
	10^{-11}$ \cite{NA62_2021}. One may establish the following rough $1\sigma $ range for the ratio in Eq. (\ref{BrKp})
	\begin{equation}
		\begin{split}
		\left( \frac{L^+}{L^+_{\text{SM}}}%
			\right)=1.26\pm 0.51,
		\end{split}
		\label{sigma}
	\end{equation}
	which has a significant uncertainty due to considerable experimental errors for the
	branching ratio. However, it may still set constraints on VLQ-extensions
	of the SM, as is the case of the $s_{14}$ dominance limit. For our model, it
	seems that larger values of $s_{14}$ are favoured and smaller values for $%
	m_{T}$ disfavoured, as can be seen from the plots in figure \ref{BrKp_plot}%
	. We consider a $95\%$ CL region where $s_{14}\in [0.03,0.05]$.
	
	\subsection*{Evaluation of $\epsilon ^{\prime }/\epsilon $}
	
	The parameter $\epsilon ^{\prime }/\epsilon $ measures direct CP violation
	in $K_L\rightarrow \pi \pi $ decays. The SM contribution can be described by
	the simplified expression \cite{Buras:2001_book}
	\[
	\left( \frac{\epsilon ^{\prime }}{\epsilon }\right) _{\text{SM}}\simeq
	F(x_{t})\text{Im}(\lambda _{t}^{K}) 
	\]
	with
	\begin{equation}
		F(x_{i})=P_{0}+P_{X}X(x_{i})+P_{Y}Y(x_{i})+P_{Z}Z(x_{i})+P_{E}E(x_{i}),
		\label{F_XYZE}
	\end{equation}
	where the Inami-Lim functions and the associated constants are detailed in
	Appendix \ref{Inami}.
	
	In a similar fashion as was done in the previous subsection, we will now
	estimate the NP contribution, with
	
	\begin{equation}
		\left( \frac{\epsilon ^{\prime }}{\epsilon }\right) _{\text{NP}}\simeq
		F(x_{T})\ \text{Im}(\lambda _{T}^{K})+(P_{X}+P_{Y}+P_{Z})\ \text{Im}(A_{ds}),
	\end{equation}
	where the second term accounts for the decoupling associated with the EW
	penguin diagrams from which the Inami-Lim functions $X(x_{i}),Y(x_{i})$ and $%
	Z(x_{i})$ are obtained \cite{Buchalla:1990qz}. In this expression, we assume
	that the constants present in $F(x_{t})$ and $F(x_{T})$ have the same values.
	
	Using Eq. (\ref{Ads}) and $c_{14},c_{24}\simeq 1$, one can write
	\begin{equation}
		\left( \frac{\epsilon ^{\prime }}{\epsilon }\right)^{\text{NP}}\simeq %
		\Tilde{F}(x_i)\text{Im}(\lambda _{T}^{K})\simeq -\Tilde{F}(x_i)%
		c_{12}^{2}s_{14}s_{24}\sin \delta ^{\prime },
	\end{equation}
	where $\Tilde{F}(x_T)\equiv F(x_{T})-\frac{x_{T}}{8}\left(
	P_{X}+P_{Y}+P_{Z}\right) $ evolves logarithmically with $x_{T}$. Once more
	it obvious that in the strict $s_{14}$ dominance limit there is no NP
	contribution.
	
	For $s_{24}\neq0$, one may use \cite{Aebischer:2020jto}
	
	\begin{equation}
		-4\times 10^{-4}\lesssim \left( \frac{\epsilon ^{\prime }}{\epsilon }\right)
		^{\text{NP}}\lesssim 10\times 10^{-4},  \label{ep_e}
	\end{equation}
	as a rough $1\sigma $ range for $\left( \epsilon ^{\prime }/\epsilon \right)
	_{\text{NP}}$. Taking into account that $\sin \delta ^{\prime }>0$ is needed
	to solve the $\epsilon _{K}$ problem, one can easily fulfil the condition in
	Eq. (\ref{ep_e}) for $s_{24}\lesssim 7.5\times 10^{-4}$ in the mass range $%
	m_{T}\in [0.685,15]$ TeV, with this allowed range becoming larger as $s_{24}$
	decreases. Therefore, the realistic $s_{14}$ dominance limit should be safe
	with regard to $\left( \epsilon ^{\prime }/\epsilon \right) $.
	
	\subsection*{Global Analysis\label{global_analysis}}
	
	Finally, we find it instructive to present a global analysis of the most
	relevant phenomenological restrictions of parameter space which apply to our 
	$s_{14}-$dominance model, in particular, the allowed parameter range for $%
	s_{14},s_{24},\delta ^{\prime }$ and $m_{T}$.
	
\begin{figure*}
	\centering
\resizebox{1.\textwidth}{!}{\includegraphics{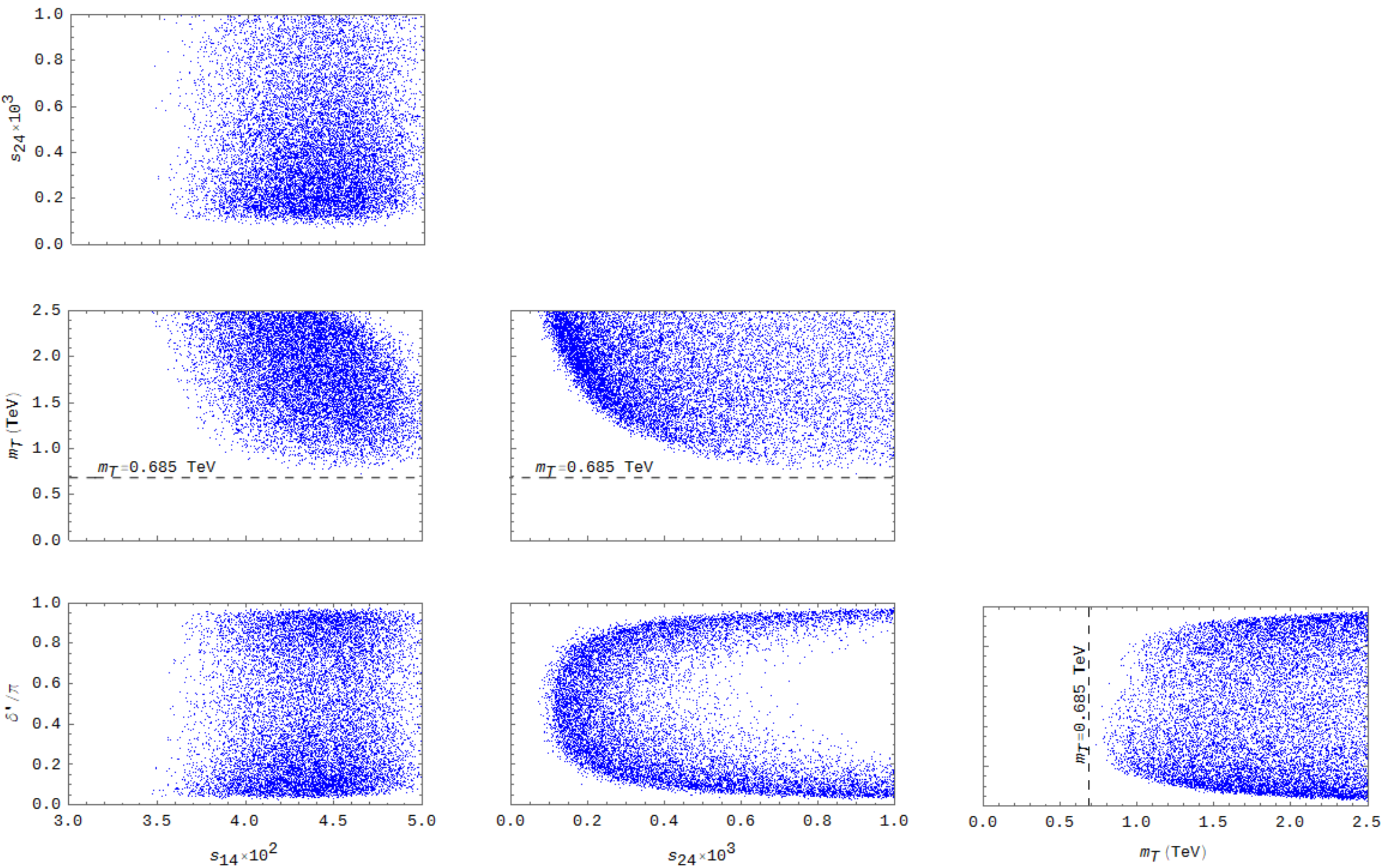}
}
	\caption{Results for the allowed parameter regions of our model verifying the conditions in Eqs. (\ref{ranges}, \ref{constraints}) and $\sqrt{\chi^2}<3$.}
	\label{simulation}     
\end{figure*}

	In figure \ref{simulation}, we present several slice-projections of the
	allowed parameter region combining the most important parameters. The values
	for $s_{14}$ are in accordance with the solution proposed for the CKM
	unitarity problem, and $s_{24},s_{34}$ are within our assumptions for $%
	s_{14} $-dominance. More concretely the parameter range for these parameters
	are
	\begin{equation}
		\begin{array}{c}
			m_{T}\in [0.685,2.5]\ \text{TeV}, \\
			\\
			s_{14}\in [0.03,0.05], \\
			\\
			s_{24},s_{34}\in
			[0,0.001],\\
			\\
			\delta _{14},\delta _{24}\in [0 ,2\pi ],
		\end{array}
		\label{ranges}
	\end{equation}
	and we impose the constraint
	\begin{equation}
	    \Delta m_{K}^{\text{NP}}<\Delta m_{K}^{\text{exp}}
	    \label{constraints}
	\end{equation}
	on the model. We also look for regions that may be accessible to upcoming generations of accelerators and therefore restrict ourselves to the study of models with masses lower than $m_T=2.5$ TeV. This is in agreement with the upper-bound presented in \cite{Belfatto:2021jhf} for models with an heavy-top where $\vert \mathcal{V}_{41}\vert\simeq 0.04$.  
	
	The points displayed in figure \ref{simulation} correspond to points that not only verify Eq. (\ref{constraints}) but also deviate less than $3\sigma$ from current experimental data, with $n\sigma$ defined as $n\sigma=\sqrt{\chi^2}$ and
	\begin{equation}
	    \begin{split}
	        \chi^2=&\sum_{i,j}\left(\frac{|V_{ij}|-|V_{ij}|_c}{\sigma_{ij}}\right)^2+\left(\frac{\gamma-\gamma_c}{\sigma_\gamma}\right)^2\\
	        & +\left(\frac{|\varepsilon^\text{NP}_K|-|\varepsilon^\text{NP}_K|_c}{\sigma_\varepsilon}\right)^2
	        +\left(\frac{\left(L^+/L^+_\text{SM}\right)-\left(L^+/L^+_\text{SM}\right)_c}{\sigma_L}\right)^2\\
	        &
	        +\left(\frac{\left(\epsilon^\prime/\epsilon\right)^\text{NP}-\left(\epsilon^\prime/\epsilon\right)^\text{NP}_c}{\sigma_{\epsilon^\prime/\epsilon}}\right)^2
	    \end{split}
	    \label{chi2}
	\end{equation}
 where for $|V_{ij}|$ we take the most relevant moduli of the SM mixing matrix entries, given by the PDG \cite{Zyla:2020zbs}, as well as the value of the rephasing 
 invariant phase $\gamma\equiv \text{arg}(-V_{ud}V_{cb}$  $V^\ast_{ub}V^\ast_{cd})$. 
 The measurement of this quantity is associated with SM tree-level dominated $B$-meson physics and is, 
 therefore, expected to remain unaffected in a model like ours, as referred also in \cite{Zyla:2020zbs}. Taking into account the current value of $\gamma=\left(72.1^{+4.1}_{-4.5}\right)^\circ$  
 we consider a central value for $\gamma_c=72.1^\circ$  and $\sigma_\gamma=4.5^\circ$ for the standard deviation. 
 
We use a similar methodology to the one presented in \cite{Branco:2021vhs}, but now adding more terms to $\chi^2$. E.g. we include the NP contribution to $\varepsilon_K$ and the new insights discussed in Eq. (\ref{eK}), with regions of parameter space where $|\epsilon^\text{NP}_K|\leq |\epsilon^\text{exp}_K|-|\epsilon^\text{SM}_K|\simeq 6.8\times 10^{-5}$. We also take into account the NP contributions associated to the decay $K^+\rightarrow \pi^+ \overline{\nu}\nu$ and the parameter $\varepsilon^\prime/\varepsilon$. The constraints set by these observables lead to the lower-bound for a heavy-top mass of around $m_T\approx 800$ GeV apparent from figure \ref{simulation}. Additionally, the kaon decay in particular restricts the allowed range of $s_{14}$ to roughly $s_{14}\in [0.035, 0.050]$ as figure \ref{BrKp_plot} previously suggested.

Note that we do not include constraints associated with other observables, such as $%
	\Delta m_{B_{d,s}}$ and $x_{D}$, because, as it was shown, their NP
	contributions are extremely suppressed in the limit of $s_{14}$ dominance.
	Furthermore, plots involving $s_{34}$ are omitted as, within the range in Eq. (\ref{ranges}), there is no noticeable influence of importance on the
	outcome of the allowed parameter region. 
	
	In the Example II of Appendix \ref{numex} we present a numerical case with a mass $m_T=1477$ GeV for the extra heavy up-quark and 
	\begin{equation}
	    \begin{array}{lllll}
	      \theta_{12}= 0.22579, & & \theta_{13}= 0.0038275, & & \theta_{23}= 0.039524, \\
	      \\
	      \theta_{14}= 0.045334, & & \theta_{24}= 7.412\times 10^{-4}, & &  \theta_{34}= 2.346\times 10^{-4}, \\
	      \\
	     \delta= 0.382\pi, & & \delta_{14}= 1.872 \pi, & & \delta_{24}=1.979\pi. \\
	    \end{array}
	\end{equation}
	leading to $\sqrt{\chi^2}\simeq 2.25$.

	\section{Conclusions}
	
	We have shown that there is a minimal extension of the SM involving the
	introduction of an up-type vector-like quark $T$, which provides a simple
	solution to the CKM unitarity problem. The heavy quark $T$ decays dominantly
	to light quarks, in contrast with the usual assumption that $T$ decays
	predominantly to the $b$ quark. Therefore, these unusual $T$ decay patterns
	should be taken into account in the experimental search for vector-like
	quarks. We have adopted the Botella-Chau parametrization of a $4\times 4$
	unitary matrix which, in contrast to the PDG parametrization, has three more
	angles $s_{14}$, $s_{24}$ and $s_{34}$ and two extra phases.
	
	We have shown that New Physics contributions e.g. to $K^0-\overline{K}^0$
	and $B_{d,s}^0-\overline{B}_{d,s}^0$ mixing or in the decays $%
	K_{L}\rightarrow \pi ^0\overline{\nu }\nu $, $K^{+}\rightarrow \pi ^{+}%
	\overline{\nu }\nu $ and new contributions to $\epsilon ^{\prime }/\epsilon $
	can be well within the limits of EWPM's.
	
	We have also used a recently introduced upper-bound on $|\epsilon^
		\text{NP}_{K}|$, which severely restricts various models, to test our own model
	with exact $s_{14}$ dominance.
	
	We have pointed out that, in the limit of exact $s_{14}$ dominance, the new
	contribution to $\epsilon _{K}$ is too large. When this limit is relaxed,
	allowing for a non-vanishing angle $s_{24}$, we then show that the leading
	order terms of $|\epsilon^{\text{NP}}_{K}|$ can be expressed as the sum of
	terms proportional to the usual CP-violating PDG phase $\delta $ and terms
	that are proportional to a new phase $\delta ^{\prime }=\delta _{24}-\delta
	_{14}$, i.e. to the difference of the other two phases of the BC
	parametrization. One can then check that there exists a reasonable parameter
	region, where these two terms may cancel each other, and that allows for the
	mass of the $T$ quark to vary between around $800$ GeV and $2.5$ TeV.
	Thus, we find that the New Physics contribution to $\epsilon _{K}$ can be
	agreement with the set upper-bound, and therefore with experiment, without
	changing the main predictions of the model, in particular the predicted
	pattern of $T$ decays.
	
	\appendix
	\section{ Numerical Examples\label{numex}}

	To stress and exemplify the claims made here, we give, in this Appendix, two
	exact numerical examples.
	
	\subsection*{Example I: Absolute dominance}
	
	As an example of exact $s_{14}$ dominance, consider the following up-sector
	mass matrix
	
	\begin{equation}
	    \small
		\mathcal{M}_{u}=\left(
        \begin{array}{cccc}
         0 & 0 & 0 & 65.2612 \\
         0 & 0 & 7.09671 & 14.848 e^{1.94715 i} \\
         0 & 19.3662 & 172.739 & 3.82017 e^{-1.5659 i} \\
         0.0397187 & 1.63395 & 32.6789 e^{-1.51428 i} & 1475.32 \\
        \end{array}
        \right),
	\end{equation}
	given in GeV at the $M_{Z}$ scale. The up-type quark masses are then, at
	this scale,
	
	\begin{equation}
		\begin{array}{cc}
			m_{u}=0.0018\text{ GeV}, & m_{c}=0.77\text{ GeV}, \\
			\\
			m_{t}=174 \text{ GeV}, & m_T=1477\text{ GeV}.
		\end{array}
		\label{spect}
	\end{equation}
	
	In the basis where the down sector mass matrix is diagonal, the matrix $%
	\mathcal{V}^{\dagger }$ which diagonalizes $\mathcal{M}_{u}$ on the left
	will have absolute value
	\begin{equation}
		\small
		|\mathcal{V}^{\dagger }|\simeq \left(
\begin{array}{cccc}
 0.973609 & 0.223644 & 0.00382359 & 0.0453188 \\
 0.223754 & 0.973844 & 0.0395133 & 0. \\
 0.008257 & 0.03883 & 0.999212 & 0. \\
 0.0441681 & 0.0101457 & 0.000173458 & 0.998973 \\
\end{array}
\right)
	\end{equation}
	Recall that $\mathcal{V}^{CKM}$ is given by a $4\times 3$ matrix of the first three
	columns of this matrix.
	
	We obtain also for the rephasing invariant phases 
	\begin{equation}
		\begin{array}{c}
			\sin (2\beta )\equiv \sin \left[ 2 \ \text{arg}\left(-V_{cd}V_{tb}V^\ast_{cb}V^\ast_{td}\right) \right]\simeq 0.764, \\
			\\
			\gamma \equiv \text{arg}\left(-V_{ud}V_{cb}V^\ast_{ub}V^\ast_{cd}\right)\simeq 68.7^{\circ }, \\
			\\
			\beta_s \equiv \text{arg}\left(-V_{cb}V_{ts}V^\ast_{cs}V^\ast_{tb}\right)\simeq 0.0206, \\
			\\
			\beta_K\equiv \text{arg}\left(-V_{us}V_{cd}V^\ast_{ud}V^\ast_{cs}\right)\simeq 6.464\times  10^{-4}, \\
		\end{array}
		\label{q-bb}
	\end{equation}
	and the CP-violation invariant, defined as 
	\begin{equation}
		J\equiv \text{Im}\left( V_{us}V_{cb}V_{ub}^{*}V_{cs}^{*}\right) ,
	\end{equation}
	has absolute value $|J|=3.070\times 10^{-5}$.
	
	For the EWPMs related quantities discussed above, we obtain the following NP contributions 
	\begin{equation}
		\begin{array}{c}
			\Delta m_{B_{d}}^{\text{NP}}\simeq 1.726\times 10^{-12}\hspace{1mm}\text{MeV}%
			,\\
			\\
			\Delta m_{B_{s}}^{\text{NP}}\simeq 2.892\times 10^{-12}\hspace{1mm}%
			\text{MeV}, \\ 
			\\
			\Delta m_{K}^{\text{NP}}\simeq 1.192\times 10^{-13}\hspace{1mm}\text{MeV}, \\
			\\
			|\epsilon^{\text{NP}}_{K}|\ \simeq 5.889\times 10^{-3},
		\end{array}
		\label{ewpm1}
	\end{equation}
	which, as stated, clearly emphasises the problem with the limit $%
	s_{24}=s_{34}=0$ and the value for the parameter $|\epsilon _{K}|$.
	
	\subsection*{Example II: Realistic dominance with very small $s_{24},s_{34}$}
	\label{exII}
	
	To exemplify a more realistic case near to our exact $s_{14}$ dominance, but
	with very small $s_{24},s_{34}$, we now consider a slightly different
	up-mass matrix (in $GeV$ at the $M_{Z}$ scale) 
	\begin{equation}
	    \small
		\mathcal{M}_{u}=\left(
        \begin{array}{cccc}
         0 & 0 & 0 & 65.033 \\
         0 & 0 & 7.12124 & 15.8436 e^{1.92462 i} \\
         0 & 19.3672 & 172.73 & 4.21828 e^{-1.56762 i} \\
         0.0397187 & 1.63403 & 32.7938 e^{-1.51551 i} & 1475.32 \\
        \end{array}
        \right),
	\end{equation}
	which leads to the same mass spectrum as the one in Eq. (\ref{spect}) and to 
	\begin{equation}
	\small
		|\mathcal{V}^{\dagger }|\simeq\left(
        \begin{array}{cccc}
         0.973609 & 0.223644 & 0.00382359 & 0.0453188 \\
         0.223785 & 0.973837 & 0.0395133 & 0.000740405 \\
         0.00824668 & 0.0388324 & 0.999212 & 0.000234355 \\
         0.0440136 & 0.010821 & 0.000312045 & 0.998972 \\
        \end{array}
        \right).
	\end{equation}
	
	The rephasing invariant phases are very similar 
	\begin{equation}
		\begin{array}{cc}
			\sin (2\beta )\simeq 0.764, & \gamma \simeq 68.7^{\circ }, \\
			\\
			\beta_s \simeq 0.0206, & \beta_K\simeq 5.950\times 10^{-4},
		\end{array}
		\label{q-bbb}
	\end{equation}
	as is the CP-violating invariant $|J|=3.070\times 10^{-5}$.
	
	The observables associated with the EWPMs have the following NP contributions 
	\begin{equation}
		\begin{array}{c}
			\Delta m_{B_{d}}^{\text{NP}}\simeq 3.119\times 10^{-12}\hspace{1mm}\text{MeV}%
			, \\
			\\
			\Delta m_{B_{s}}^{\text{NP}}\simeq 5.547\times 10^{-12}\hspace{1mm}
			\text{MeV}, \\ 
			\\
			\Delta m_{K}^{\text{NP}}\simeq 1.356\times 10^{-12}\hspace{1mm}\text{MeV},\\
			\\
			 |\epsilon^{\text{NP}}_{K}|\ \simeq 6.592\times 10^{-5},
		\end{array}
		\label{ewpm2}
	\end{equation}
	
	As it is clear, the problem with $\epsilon _{K}$ is now successfully solved.
	Comparing Eq. (\ref{ewpm1}) and Eq. (\ref{ewpm2}) one also sees that
	although noticeable changes to $\Delta m_{B_d}^{\text{NP}}$ and $\Delta
	m_{Bs}^{\text{NP}}$ took place, these are still small and in no way
	compromise the safety of the model.
	
	\section{Inami-Lim functions \label{Inami}
	\label{appendix_B}}

	The Inami-Lim functions used throughout this paper are given by \cite
	{Inami:1980fz,Buchalla:1990qz}
	
	\begin{equation}
	        S_{ij} \equiv S(x_{i},x_{j}) = x_{i}x_{j}\left[ \frac{\log x_{i} \left( 1-2x_{i}+\frac{x_{i}^{2}}{4}\right)}{%
			(x_{i}-x_{j})(1-x_{i})^{2}}+ (x_{i}\leftrightarrow x_{j})
		\right]-\frac{3x_{i}x_{j}}{4(1-x_{i})(1-x_{j})}, 
		\label{sij}
	\end{equation}
	
	\begin{equation}
	        S_{i}\equiv S(x_{i})\equiv \lim_{x_{j}\rightarrow x_{i}}S(x_{i},x_{j})=\frac{%
			x_{i}}{(1-x_{i})^{2}}\left( 1-\frac{11}{4}x_{i}+\frac{x_{i}^{2}}{4}\right) -%
		\frac{3}{2}\frac{x_{i}^{3}\log x_{i}}{(1-x_{i})^{3}},
		  \label{si}
	\end{equation}
	
	\begin{equation}
	    X(x_i)=\frac{x_i}{8(x_i-1)}\left(x_i+2+\frac{3x_i-6}{x_i-1}\log x_i\right),
		\label{X_IL}
	\end{equation}
	
	\begin{equation}
		Y(x_i)=\frac{x_i}{8(x_i-1)}\left(x_i-4+\frac{3x}{x_i-1}\log x_i\right),
	\end{equation}
	
	\begin{equation}
	       Z(x_i)= -\frac{\log x_i}{9}+\frac{18 x^4_i-163 x^3_i+259 x^2_i-108 x_i}{%
			144(x_i-1)^3}+\frac{32 x^4_i-38 x^3_i-15 x^2_i+18 x_i}{72(x_i-1)^4}\log x_i,
	\end{equation}
	
	\begin{equation}
	        E(x_i)=-\frac{2\log x_i}{3}+\frac{x_i (18 - 11 x_i - x_i^2)}{12(1-x_i)^3}+\frac{x_i^2 (15 - 16 x_i + 4 x_i^2)}{6(1-x_i)^4}\log x_i.
	\end{equation}
	
	All these functions are gauge invariant, however, $X(x_{i}),$ $Y(x_{i})$ and $
	Z(x_{i})$ correspond to linear combinations of gauge-dependent functions. $%
	X(x_{i})$ and $Y(x_{i})$ are obtained by combining box functions with $Z$
	penguin functions, whereas $Z(x_{i})$ is obtained by combining photon and $Z$
	penguin functions. $S(x_{i},x_{j})$ is a box diagram function that is
	relevant in meson mixings and $E(x_{i})$ is associated with gluon penguins.
	
	The function $F(x_{i})$ in Eq. (\ref{F_XYZE}), relevant to the study of $%
	\epsilon ^{\prime }/\epsilon $, is a linear combination of $%
	X(x_{i}),Y(x_{i}),Z(x_{i})$ and $E(x_{i})$. We use the following values for
	the constants entering this expression \cite{Buras:2015yba}
	
	\begin{equation}
		\begin{array}{c}
			P_{0}\simeq -3.392+15.3037\ B_{6}^{(1/2)}+1.7111\ B_{8}^{(3/2)}, \\ 
			\\ 
			P_{X}\simeq 0.655+0.02902\ B_{6}^{(1/2)}, \\ 
			\\ 
			P_{Y}\simeq 0.451+0.1141\ B_{6}^{(1/2)}, \\ 
			\\ 
			P_{Z}\simeq 0.406-0.0220\ B_{6}^{(1/2)}-13.4434\ B_{8}^{(3/2)}, \\ 
			\\ 
			P_{E}\simeq 0.229-1.7612\ B_{6}^{(1/2)}+0.6525\ B_{8}^{(3/2)},
		\end{array}
		\label{Pconst}
	\end{equation}
	as well as the central values of $B_{6}^{(1/2)}=1.11\pm 0.20$ and $%
	B_{8}^{(3/2)}=0.70\pm 0.04$ \cite{Aebischer:2020jto}.
	
\begin{figure}[h]
		\centering
		\begin{tikzpicture}
			\begin{feynman}
				\vertex (a1) {\(\overline s\)};
				\vertex[right=1.9cm of a1] (a2);
				\vertex[right=1.9cm of a2] (a3);
				\vertex[right=1.6cm of a3] (a4) {\(\overline{\nu}_\ell\)};
				
				\vertex[below=1.7cm of a1] (b1) {\(d\)};
				\vertex[right=1.9cm of b1] (b2);
				\vertex[right=1.9cm of b2] (b3);
				\vertex[right=1.6cm of b3] (b4) {\(\nu_\ell \hspace{0.5mm}\)};
				
				\diagram* {
					{[edges=fermion]
						(b1) -- (b2) -- [edge label=\(u_i\)] (a2) -- (a1),
						(a4)--(a3)--[edge label=\(\ell\)](b3)--(b4) ,
					},
					(a2) -- [boson, edge label=\(W\)] (a3),
					(b2) -- [boson, edge label'=\(W\)] (b3)
				};
			\end{feynman}
		\end{tikzpicture}
		\caption{Box diagram contributing to $K_L\rightarrow \pi^0 \overline{\nu}\nu$
			and $K^+\rightarrow \pi^+ \overline{\nu}\nu$, from which $X(x_i)$ in Eq. (%
			\ref{X_IL}) is obtained, with $u_{i,j}=u,c,t,T$ and $\ell=e,\mu,\tau$.}
		\label{leptonbox}
	\end{figure}
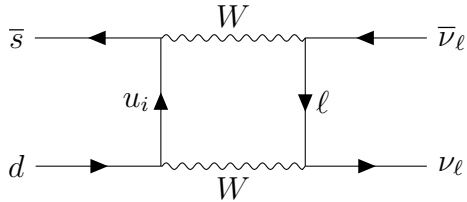

	The correction $X^{\text{NNL}}(x_{c})$ used in Eq. (\ref{BrKp}) is important
	because, as mentioned above, the Inami-Lim function $X(x_{i})$ is obtained
	from combining the contributions of penguin and box diagrams to neutrino
	decays of mesons. For the kaon case, the relevant box diagrams are the ones
	presented in figure (\ref{leptonbox}).
	
	The expression for $X(x_{i})$ in Eq. (\ref{X_IL}) is obtained by taking the
	limit of vanishing masses for the leptons involved in the loop so that this
	function involves solely the mass of the up-type quark running inside the
	loop. This is a good approximation for the top and heavy top contributions
	given that $m_{t},m_{T}\gg m_{\tau }$, however for the charm quark one has $%
	m_{c}<m_{\tau }$ and Eq. (\ref{X_IL}) is no longer valid. Hence, it should
	be replaced by
	
	\begin{equation}
		X^{\text{NNL}}(x_{c})=X_{\text{SD}}^{\text{NNL}}(x_{c})+\delta X(x_{c}),
	\end{equation}
	where $\delta X(x_{c})$ is the long-distance contribution. The
	short-distance piece is, at NNLO, given by
	
	\begin{equation}
		X_{\text{SD}}^{\text{NNL}}(x_{c})=\frac{2}{3}X_{e}^{\text{NNL}}(x_{c})+\frac{%
			1}{3}X_{\tau }^{\text{NNL}}(x_{c}),
	\end{equation}
	so that the contributions involving the lepton $\tau $ and the remaining
	lighter leptons are considered separately. Following \cite{Buras_2015} one
	can approximate this quantity with $X^{\text{NNL}}(x_{c})\simeq 1.04\times
	10^{-3}$.
	
	\section*{Acknowledgments}
	
	
	This work was partially supported by Funda{\c{c}}{\~{a}}o para a Ci{\^{e}}%
	ncia e a Tecnologia (FCT, Portugal) through the projects CFTP-FCT Unit 777
	(UIDB/00777/2020 and UIDP/00777/2020), PTDC/FIS-PAR/29436/2017, and
	CERN/FIS-PAR/0008/2019, which are partially funded through POCTI (FEDER),
	COMPETE, QREN and EU. G.C.B.~and M.N.R.~benefited from discussions that were
	prompted through the HARMONIA project of the National Science Centre,
	Poland, under contract UMO-2015/18/M/ST2/00518 (2016-2019), which has been
	extended. F.J.B. research was founded by the Spanish grant
	PID2019-106448GB-C33 (AEI/FEDER, UE) and by Generalitat Valenciana, under
	grant PROMETEO 2019-113.
	
	
	\providecommand{\noopsort}[1]{}\providecommand{\singleletter}[1]{#1}%
	
	\providecommand{\href}[2]{#2}\begingroup\raggedright\endgroup
		

\begin{thebibliography}{70}
%
%
\bibitem{Seng:2018yzq}
			C.-Y. Seng, M.~Gorchtein, H.~H. Patel and M.~J. Ramsey-Musolf, \emph{{Reduced
					Hadronic Uncertainty in the Determination of $V_{ud}$}},
			\href{https://doi.org/10.1103/PhysRevLett.121.241804}{\emph{Phys. Rev. Lett.} {\bfseries 121} (2018) 241804} 
			[\href{https://arxiv.org/abs/1807.10197}{{\ttfamily 1807.10197}}].
			
			\bibitem{Seng:2018qru}
			C.~Y. Seng, M.~Gorchtein and M.~J. Ramsey-Musolf, \emph{{Dispersive evaluation
					of the inner radiative correction in neutron and nuclear $\beta$ decay}},
			\href{https://doi.org/10.1103/PhysRevD.100.013001}{\emph{Phys. Rev. D} {\bfseries 100} (2019) 013001}
			[\href{https://arxiv.org/abs/1812.03352}{{\ttfamily 1812.03352}}].
			
			\bibitem{Czarnecki:2019mwq}
			A.~Czarnecki, W.~J. Marciano and A.~Sirlin, \emph{{Radiative Corrections to
					Neutron and Nuclear Beta Decays Revisited}},
			\href{https://doi.org/10.1103/PhysRevD.100.073008}{\emph{Phys. Rev. D} {\bfseries 100} (2019) 073008}
			[\href{https://arxiv.org/abs/1907.06737}{{\ttfamily 1907.06737}}].
			
			
			\bibitem{Seng:2020wjq}
			C.-Y. Seng, X.~Feng, M.~Gorchtein and L.-C. Jin, \emph{{Joint lattice
					QCD\textendash{}dispersion theory analysis confirms the quark-mixing top-row
					unitarity deficit}},
			\href{https://doi.org/10.1103/PhysRevD.101.111301}{\emph{Phys. Rev. D} {\bfseries 101} (2020) 111301}
			[\href{https://arxiv.org/abs/2003.11264}{{\ttfamily 2003.11264}}].
			
			\bibitem{Hayen:2020cxh}
			L.~Hayen, \emph{{Standard Model $\mathcal{O}(\alpha)$ renormalization of $g_A$
					and its impact on new physics searches}},
			\href{https://doi.org/10.1103/PhysRevD.103.113001}{\emph{Phys. Rev. D} {\bfseries 103} (2021) 113001}
			\href{https://arxiv.org/abs/2010.07262}{{\ttfamily 2010.07262}}.
			
			
			\bibitem{Shiells:2020fqp}
			K.~Shiells, P.~G. Blunden and W.~Melnitchouk, \emph{{Electroweak axial
					structure functions and improved extraction of the $V_{ud}$ CKM matrix
					element}}, \href{https://doi.org/10.1103/PhysRevD.104.033003}{\emph{Phys. Rev. D} {\bfseries 104} (2021) 033003}
			\href{https://arxiv.org/abs/2012.01580}{{\ttfamily 2012.01580}}.
			
			\bibitem{Czarnecki:2004cw}
			A.~Czarnecki, W.~J. Marciano and A.~Sirlin, \emph{{Precision measurements and
					CKM unitarity}},
			\href{https://doi.org/10.1103/PhysRevD.70.093006}{\emph{Phys. Rev. D} {\bfseries 70} (2004) 093006}
			[\href{https://arxiv.org/abs/hep-ph/0406324}{{\ttfamily hep-ph/0406324}}].
			
			
			\bibitem{Coutinho:2019aiy}
			Antonio M. Coutinho, Andrea Crivellin, \emph{{Global Fit to Modified Neutrino Couplings}},
			\href{https://doi.org/10.1103/PhysRevLett.125.071802}{\emph{Phys. Rev. Lett.} {\bfseries 125} (2020) 071802}
			[\href{https://arxiv.org/abs/1912.08823}{{\ttfamily 1912.08823}}].
			
			\bibitem{Aoki:2021kgd}
			Y. Aoki {\it et al}, \emph{{FLAG Review 2021}},
			[\href{https://arxiv.org/abs/2111.09849}{{\ttfamily 2111.09849}}].
			
			\bibitem{Belfatto:2019swo}
			B.~Belfatto, R.~Beradze and Z.~Berezhiani, \emph{{The CKM unitarity problem: A
					trace of new physics at the TeV scale?}},
			\href{https://doi.org/10.1140/epjc/s10052-020-7691-6}{\emph{Eur. Phys. J. C} {\bfseries 80} (2020) 149}
			[\href{https://arxiv.org/abs/1906.02714}{{\ttfamily 1906.02714}}].
			
			\bibitem{Branco:2021vhs}
			G. C. Branco, J. T. Penedo, Pedro M. F. Pereira, M. N. Rebelo and J. I. Silva-Marcos, 
			\href{https://doi.org/10.1007/JHEP07(2021)099}{\emph{{Addressing the CKM unitarity problem with a vector-like up quark}}, 
				JHEP \textbf{07} (2021), 099}
			[ \href{https://arxiv.org/abs/2103.13409}{{\ttfamily 2103.13409}}].
			
			\bibitem{Belfatto:2021jhf}
			B.~Belfatto and Z.~Berezhiani,
			\href{https://doi.org/10.1007/JHEP10(2021)079}{\emph{{Are the CKM anomalies induced by vector-like quarks? Limits from flavor changing and Standard Model precision tests}}, 
				JHEP \textbf{10} (2021), 079}
			[\href{https://arXiv.org/abs/2103.05549}{{\ttfamily  2103.13409}}].
			
			\bibitem{Crivellin:2021bkd}
			A.~Crivellin, M.~Hoferichter, M.~Kirk, C.~A.~Manzari and L.~Schnell,
			\href{https://doi.org/10.1007/JHEP10(2021)221}{\emph{{First-generation new physics in simplified models: from low-energy parity violation to the LHC}},
				JHEP \textbf{10} (2021), 221}
			[\href{https://arXiv.org/abs/2107.13569}{{\ttfamily  2107.13569}}].
			
			\bibitem{Bento:1991ez}
			L.~Bento, G.~C. Branco and P.~A. Parada, \emph{{A Minimal model with natural
					suppression of strong CP violation}},
			\href{https://doi.org/10.1016/0370-2693(91)90530-4} {\emph{Phys. Lett. B} {\bfseries 267} (1991) 95}.
			
			\bibitem{Nardi:1991rg}
			E.~Nardi, E.~Roulet and D.~Tommasini, \emph{{Global analysis of fermion mixing with exotics}},
			\href{https://doi.org/10.1016/0550-3213(92)90566-T}{\emph{Nucl. Phys. B}  {\bfseries 386} (1992) 239}.
			
			
			\bibitem{Branco:1992wr}
			G.~C. Branco, T.~Morozumi, P.~A. Parada and M.~N. Rebelo, \emph{{$CP$
					asymmetries in $B^0$ decays in the presence of flavor-changing neutral
					currents}}, \href{https://doi.org/10.1103/PhysRevD.48.1167}{\emph{Phys. Rev.}
				{\bfseries D48} (1993) 1167}
			
			\bibitem{Branco:1992uy}
			G.~C. Branco, P.~A. Parada, T.~Morozumi and M.~N. Rebelo, \emph{{Effect of
					flavor changing neutral currents in the leptonic asymmetry in B(d) decays}},
			\href{https://doi.org/10.1016/0370-2693(93)90098-3}{\emph{Phys. Lett. B} {\bfseries 306} (1993) 398},
			
			\bibitem{delAguila:1997vn}
			F.~del Aguila, J.~A. Aguilar-Saavedra and G.~C. Branco, \emph{{CP violation
					from new quarks in the chiral limit}},
			\href{https://doi.org/10.1016/S0550-3213(97)00708-6}{\emph{Nucl. Phys. B} {\bfseries 510} (1998) 39},
			[\href{https://arxiv.org/abs/hep-ph/9703410}{{\ttfamily hep-ph/9703410}}].
			
			\bibitem{Barenboim:1997pf}
			G.~Barenboim and F.~J. Botella, \emph{{Delta F=2 effective Lagrangian in
					theories with vector - like fermions}},
			\href{https://doi.org/10.1016/S0370-2693(98)00695-9}{\emph{Phys. Lett. B}{\bfseries 433} (1998) 385}
			[\href{https://arxiv.org/abs/hep-ph/9708209}{{\ttfamily hep-ph/9708209}}].
			
			\bibitem{Barenboim:1997qx}
			G.~Barenboim, F.~J. Botella, G.~C. Branco and O.~Vives, \emph{{How sensitive to
					FCNC can B0 CP asymmetries be?}},
			\href{https://doi.org/10.1016/S0370-2693(97)01515-3}{\emph{Phys. Lett. B}
				{\bfseries 422} (1998) 277}
			[\href{https://arxiv.org/abs/hep-ph/9709369}{{\ttfamily hep-ph/9709369}}].
			
			\bibitem{delAguila:2000aa}
			F.~del Aguila, M.~Perez-Victoria and J.~Santiago, \emph{{Effective description
					of quark mixing}},
			\href{https://doi.org/10.1016/S0370-2693(00)01071-6}{\emph{Phys. Lett. B}
				{\bfseries 492} (2000) 98}
			[\href{https://arxiv.org/abs/hep-ph/0007160}{{\ttfamily hep-ph/0007160}}].
			
			\bibitem{delAguila:2000rc}
			F.~del Aguila, M.~Perez-Victoria and J.~Santiago, \emph{{Observable
					contributions of new exotic quarks to quark mixing}},
			\href{https://doi.org/10.1088/1126-6708/2000/09/011}{\emph{JHEP} {\bfseries 09} (2000) 011} 
			[\href{https://arxiv.org/abs/hep-ph/0007316}{{\ttfamily hep-ph/0007316}}].
			
			\bibitem{Botella:1985gb}
			F.~Botella and L.-L. Chau, \emph{{Anticipating the Higher Generations of Quarks
					from Rephasing Invariance of the Mixing Matrix}},
			\href{https://doi.org/10.1016/0370-2693(86)91468-1}{\emph{Phys. Lett. B}
				{\bfseries 168} (1986) 97}.
			
			
			\bibitem{Brod_2012}
			J.~Brod and M.~Gorbahn,
			\emph{{Next-to-Next-to-Leading-Order Charm-Quark Contribution to the $CP$ Violation Parameter $\epsilon_K$ and $\Delta M_K$}},
			\href{https://doi.org/10.1103/PhysRevLett.108.121801}{Phys. Rev. Lett. \textbf{108} (2012), 121801}
			[\href{https://arxiv.org/abs/1108.2036} {arXiv:1108.2036}].
			
			\bibitem{Zyla:2020zbs}
			{\scshape Particle Data Group} collaboration, P.~Zyla et~al., \emph{{Review of Particle Physics}}, 
			\href{https://doi.org/10.1093/ptep/ptaa104}{\emph{PTEP} {\bfseries 2020} (2020) 083C01}.
			
			
			\bibitem{Sirunyan_2018}
			A.~M.~Sirunyan \textit{et al.} [CMS],
			\emph{{Search for vectorlike light-flavor quark partners in proton-proton collisions at $\sqrt s$ =8  TeV}},
			\href{https://doi.org/10.1103/PhysRevD.97.072008}{Phys. Rev. D \textbf{97} (2018), 072008}
			[\href{https://arxiv.org/abs/1708.02510}{arXiv:1708.02510} ].
			
			
			\bibitem{Cacciapaglia_2012}
			G.~Cacciapaglia, A.~Deandrea, L.~Panizzi, N.~Gaur, D.~Harada and Y.~Okada,
			\emph{{Heavy Vector-like Top Partners at the LHC and flavour constraints}},
			\href{https://doi.org/10.1007/JHEP03(2012)070}{JHEP \textbf{03} (2012), 070}
			[\href{https://arxiv.org/abs/1108.6329}{arXiv:1108.6329}].
			
			\bibitem{Branco:1999fs}
			G.~C.~Branco, L.~Lavoura and J.~P.~Silva,
			\emph{{CP Violation,}}
			Int. Ser. Monogr. Phys. \textbf{103} (1999), 1-536.
			
			\bibitem{Inami:1980fz}
			T.~Inami and C.~S.~Lim,
			\emph{{Effects of Superheavy Quarks and Leptons in Low-Energy Weak Processes k(L) ---\ensuremath{>} mu anti-mu, K+ ---\ensuremath{>} pi+ Neutrino anti-neutrino and K0 \ensuremath{<}---\ensuremath{>} anti-K0}},
			\href{https://doi.org/10.1143/PTP.65.297}{Prog. Theor. Phys. \textbf{65} (1981), 297}
			[erratum: Prog. Theor. Phys. \textbf{65} (1981), 1772]
			
			\bibitem{Chetyrkin_2009}
			K.~G.~Chetyrkin, J.~H.~Kuhn, A.~Maier, P.~Maierhofer, P.~Marquard, M.~Steinhauser and C.~Sturm,
			\emph{{Charm and Bottom Quark Masses: An Update}},
			\href{https://doi.org/10.1103/PhysRevD.80.074010}{Phys. Rev. D \textbf{80} (2009), 074010}
			[\href{https://arxiv.org/abs/0907.2110}{arXiv:0907.2110}].
			
			
			\bibitem{Huang_2020}
			X.~D.~Huang, X.~G.~Wu, J.~Zeng, Q.~Yu, X.~C.~Zheng and S.~Xu,
			\emph{{Determination of the top-quark $\overline{MS}$ running mass via its perturbative relation to the on-shell mass with the help of the principle of maximum conformality}},
			\href{https://doi.org/10.1103/PhysRevD.101.114024}{Phys. Rev. D \textbf{101} (2020) no.11, 114024}
			[\href{https://arxiv.org/abs/2005.04996}{arXiv:2005.04996}].
			
			\bibitem{Buras_2010}
			A.~J.~Buras, B.~Duling, T.~Feldmann, T.~Heidsieck, C.~Promberger and S.~Recksiegel,
			\emph{{Patterns of Flavour Violation in the Presence of a Fourth Generation of Quarks and Leptons}},
			\href{https://doi.org/10.1007/JHEP09(2010)106}{JHEP \textbf{09} (2010), 106}
			[\href{https://arxiv.org/abs/1002.2126}{arXiv:1002.2126}].
			
			\bibitem{Brod_2010}
			J.~Brod and M.~Gorbahn,
			\emph{{$\epsilon_K$ at Next-to-Next-to-Leading Order: The Charm-Top-Quark Contribution}},
			\href{https://doi.org/10.1103/PhysRevD.82.094026}{Phys. Rev. D \textbf{82} (2010), 094026}
			[\href{https://arxiv.org/abs/1007.0684}{arXiv:1007.0684}].
			
			\bibitem{Bobeth_2017}
			C.~Bobeth, A.~J.~Buras, A.~Celis and M.~Jung,
			\emph{{Patterns of Flavour Violation in Models with Vector-Like Quarks}},
			\href{https://doi.org/10.1007/JHEP04(2017)079}{JHEP \textbf{04} (2017), 079}
			[\href{https://arxiv.org/abs/1609.04783}{arXiv:1609.04783}].
			
			\bibitem{Aguilar-Saavedra:2002phh}
			J.~A.~Aguilar-Saavedra,
			\emph{{Effects of mixing with quark singlets}},
			\href{https://doi.org/10.1103/PhysRevD.69.099901}{Phys. Rev. D \textbf{67} (2003), 035003},
			[erratum: Phys. Rev. D \textbf{69} (2004), 099901]
			[\href{https://arxiv.org/abs/hep-ph/0210112}{hep-ph/0210112}].
			
			
			\bibitem{Aoki_2020}
			S.~Aoki \textit{et al.} [Flavour Lattice Averaging Group],
			\emph{{FLAG Review 2019: Flavour Lattice Averaging Group (FLAG)}},
			\href{https://doi.org/10.1140/epjc/s10052-019-7354-7}{Eur. Phys. J. C \textbf{80} (2020) no.2, 113}
			[\href{https://arxiv.org/abs/1902.08191}{arXiv:1902.08191}].
			
			
			\bibitem{Buras_1998}
			A.~J.~Buras and R.~Fleischer,
			\emph{{Quark mixing, CP violation and rare decays after the top quark discovery}},
			\href{https://doi.org/10.1142/9789812812667\_0002}{Adv. Ser. Direct. High Energy Phys. \textbf{15} (1998), 65-238}
			[\href{https://arxiv.org/abs/hep-ph/9704376}{hep-ph/9704376}].
			
			\bibitem{Buras_2008}
			A.~J.~Buras and D.~Guadagnoli,
			\emph{{Correlations among new CP violating effects in $\Delta$ F = 2 observables}},
			\href{https://doi.org/10.1103/PhysRevD.78.033005}{Phys. Rev. D \textbf{78} (2008), 033005}
			[\href{https://arxiv.org/abs/0805.3887}{arXiv:0805.3887}].
			
			\bibitem{Brod_2020}
			J.~Brod, M.~Gorbahn and E.~Stamou,
			\emph{{Standard-Model Prediction of $\epsilon_K$ with Manifest Quark-Mixing Unitarity}},
			\href{https://doi.org/10.1103/PhysRevLett.125.171803}{Phys. Rev. Lett. \textbf{125} (2020) no.17, 171803}
			[\href{https://arxiv.org/abs/1911.06822}{arXiv:1911.06822}].
			
			\bibitem{Botella_2017}
			Francisco J.Botella, G.~C.~Branco, Miguel Nebot, M.~N.~Rebelo, and J.~I.~Silva-Marcos.,
			\emph{{Vector-like quarks at the origin of light quark masses and mixing}},
			\href{https://doi.org/10.1140/epjc/s10052-017-4933-3}{Eur. Phys. J. C \textbf{77} (2017)  408}
			[\href{https://arxiv.org/abs/1610.03018}{arXiv:1610.03018}]
			
			\bibitem{Balaji:2021lpr}
			Shyam Balaji,
			\emph{{Asymmetry in flavour changing electromagnetic transitions of vector-like quarks}},
			[\href{https://arxiv.org/abs/2110.05473}{arXiv:2110.05473}]
			
		
			
			\bibitem{Branco_1995}
			G.~C.~Branco, P.~A.~Parada and M.~N.~Rebelo,
			\emph{{D0 - anti-D0 mixing in the presence of isosinglet quarks}},
			\href{https://doi.org/10.1103/PhysRevD.52.4217}{Phys. Rev. D \textbf{52} (1995), 4217-4222}
			[\href{https://arxiv.org/abs/hep-ph/9501347}{hep-ph/9501347}].
			
			\bibitem{Golowich_2009}
			E.~Golowich, J.~Hewett, S.~Pakvasa and A.~A.~Petrov,
			\emph{{Relating D0-anti-D0 Mixing and D0 ---\ensuremath{>} l+ l- with New Physics}},
			\href{https://doi.org/10.1103/PhysRevD.79.114030}{Phys. Rev. D \textbf{79} (2009), 114030}
			[\href{https://arxiv.org/abs/0903.2830}{arXiv:0903.2830}].
			
			\bibitem{Buras_2010_2}
			A.~J.~Buras, B.~Duling, T.~Feldmann, T.~Heidsieck, C.~Promberger and S.~Recksiegel,
			\emph{{The Impact of a 4th Generation on Mixing and CP Violation in the Charm System}},
			\href{https://doi.org/10.1007/JHEP07(2010)094}{JHEP \textbf{07} (2010), 094}
			[\href{https://arxiv.org/abs/1004.4565}{arXiv:1004.4565}].
			
			\bibitem{Amhis_2021}
			Y.~S.~Amhis \textit{et al.} [HFLAV],
			\emph{{Averages of b-hadron, c-hadron, and $\tau $-lepton properties as of 2018}},
			\href{https://doi.org/10.1140/epjc/s10052-020-8156-7}{Eur. Phys. J. C \textbf{81} (2021) no.3, 226}
			[\href{https://arxiv.org/abs/1909.12524}{arXiv:1909.12524}].
			
			
			
			\bibitem{Aguilar-Saavedra:2004mfd}
			J.~A.~Aguilar-Saavedra,
			\emph{{Top flavor-changing neutral interactions: Theoretical expectations and experimental detection}},
			Acta Phys. Polon. B \textbf{35} (2004), 2695-2710
			[\href{https://arxiv.org/abs/hep-ph/0409342}{hep-ph/0409342}].
			
			\bibitem{2018}
			M.~Aaboud \textit{et al.} [ATLAS],
			\emph{{Search for flavour-changing neutral current top-quark decays $t\to qZ$ in proton-proton collisions at $\sqrt{s}=13$ TeV with the ATLAS detector}},
			\href{https://doi.org/10.1007/JHEP07(2018)176}{JHEP \textbf{07} (2018), 176}
			[\href{https://arxiv.org/abs/1803.09923}{arXiv:1803.09923}].
			
			\bibitem{Buchalla:1990qz}
			G.~Buchalla, A.~J.~Buras and M.~K.~Harlander,
			\emph{{Penguin box expansion: Flavor changing neutral current processes and a heavy top quark}},
			\href{https://www.sciencedirect.com/science/article/abs/pii/0550321391901862?via%3Dihub.}{Nucl. Phys. B \textbf{349} (1991), 1-47}.
			
			\bibitem{Buras:2001_book}
			Andrzej J. Buras, 
			\emph{{CP Violation and Rare Decays}},
			\emph{In Theory and Experiment Heading for New Physics}, 
			Edited by Antonino Zichichi. World Scientific (2001). ISBN 9810247931
			
			\bibitem{Buras:2015yba}
			A.~J.~Buras, M.~Gorbahn, S.~J\"ager and M.~Jamin,
			\emph{{Improved anatomy of \ensuremath{\varepsilon}'/\ensuremath{\varepsilon} in the Standard Model}},
			\href{https://doi.org/10.1007/JHEP11(2015)202}{JHEP \textbf{11} (2015), 202}
			[\href{https://arxiv.org/abs/1507.06345}{arXiv:1507.06345}].
			
			
			\bibitem{Aebischer:2020jto}
			J.~Aebischer, C.~Bobeth and A.~J.~Buras,
			\emph{{$\varepsilon '/\varepsilon $ in the Standard Model at the Dawn of the 2020s}},
			\href{https://doi.org/10.1140/epjc/s10052-020-8267-1}{Eur. Phys. J. C \textbf{80} (2020) no.8, 705}
			[\href{https://arxiv.org/abs/2005.05978}{arXiv:2005.05978}].
			
			\bibitem{Botella_2017_a}
			Francisco J. Botella, Gustavo C. Branco and Miguel Nebot,
			\emph{{Singlet Heavy Fermions as the Origin of B Anomalies in Flavour Changing Neutral Currents}},
			[\href{https://arxiv.org/abs/1712.04470}{arXiv:1712.04470}].
			
			\bibitem{Nardi_1996}
			Enrico Nardi, 
			\emph{{Top - charm flavor changing contributions to the effective bsZ vertex}},
			\href{https://doi.org/10.1016/0370-2693(95)01308-3}{Phys. Lett. B \textbf{365} (1996) 327}
			[\href{https://arxiv.org/abs/hep-ph/9509233}{hep-ph/9509233}]
			
			\bibitem{Vysotsky_2007}
			M.I. Vysotsky, 
			\emph{{New (virtual) physics in the era of the LHC}},
			\href{https://doi.org/10.1016/j.physletb.2006.12.003}{Phys. Lett. B \textbf{644} (2007) 352}
			[\href{https://arxiv.org/abs/hep-ph/0610368}{hep-ph/0610368}]
			
			\bibitem{Kopnin_2008}
			P. Kopnin and M. Vysotsky,
			\emph{{Manifestation of a singlet heavy up-type quark in the
					branching ratios of rare decays $K\rightarrow \pi \nu \overline{\nu}$, $B\rightarrow \pi \nu \overline{\nu}$ and $B\rightarrow K \nu \overline{\nu}$}},
			\href{https://doi.org/10.1134/S0021364008100019}{JETP Lett. \textbf{87} (2008) 517}
			[\href{https://arxiv.org/abs/0804.0912}{hep-ph/08040912}]
			
			\bibitem{Picek_2008}
			I. Picek and B. Radovcic,
			\emph{{Nondecoupling of terascale isosinglet quark and rare K and B decays}},
			\href{https://doi.org/10.1103/PhysRevD.78.015014}{Phys. Rev. D \textbf{78} (2008) 015014}
			[\href{https://arxiv.org/abs/0804.2216}{arXiv:0804.2216}].
			
			\bibitem{Buras_2015}
			Andrzej J. Buras, Dario Buttazzo, Jennifer Girrbach-Noe, Robert Knegjens
			\emph{{$K^+\rightarrow \pi^+ \nu \overline{\nu}$ and $K\rightarrow \pi \nu \overline{\nu}$ in the Standard Model: Status and Perspectives}},
			\href{https://doi.org/10.1007/JHEP11(2015)033}{JHEP \textbf{1511} (2015) 033}
			[\href{https://arxiv.org/abs/1503.02693}{arXiv:1503.02693}].
			
			\bibitem{NA62_2021}
			The NA62 collaboration, E. Cortina Gil et al,
			\emph{{Measurement of the very rare $K^+\rightarrow \pi^+ \nu \overline{\nu}$ decay}},
			\href{https://doi.org/10.1007/JHEP06(2021)093}{JHEP \textbf{06} (2021) 093}
			[\href{https://arxiv.org/abs/2103.15389}{arXiv:2103.15389}].
\end{thebibliography}
	\end{document}